\documentclass[11pt,a4paper]{article}
\usepackage{jcappub}
\usepackage{bm}
\usepackage{amsmath}
\usepackage{mathrsfs}
\usepackage[mathscr]{euscript}
\usepackage{feynmp}
\usepackage{array}
\usepackage{natbib}
\usepackage{ulem}
\usepackage{comment}
\usepackage{booktabs}
\usepackage{xcolor}

\usepackage{xcolor}

\def\0{{(0)}}
\def\sig0{\dot{\sigma}_0}

\def\ph0{\dot{\phi}_0}

\title{
Cosmic Birefringence from\\ Monodromic Axion Dark Energy}
\author[a,b]{Silvia Gasparotto,}
\author[a]{Ippei Obata}
\affiliation[a]{Max-Planck-Institut f{\"u}r Astrophysik, Karl-Schwarzschild-Str. 1, 85748 Garching, Germany}
\affiliation[b]{Institut de Fisica d’Altes Energies (IFAE), The Barcelona Institute of Science and Technology,
Campus UAB, 08193 Bellaterra (Barcelona), Spain}
\emailAdd{sgasparotto@ifae.es}
\emailAdd{obata@mpa-garching.mpg.de}

\abstract{
The recently reported non-zero isotropic birefringence angle in Planck 2018 polarization data provides a tantalizing hint for new physics of axions. In this paper, we explain this by a string theory motivated axion with a monodromy potential that plays the role of dark energy.
Upon using the birefringence measurement and the constraint on the equation of state for dark energy in this scenario, we find an upper bound on the axion decay constant as $f_a \lesssim 10^{16}$ GeV. This naturally gives an energy scale of order GUT and can resolve the theoretical issue of super-Planckian field range of the conventional axion dark energy model. We further study the implications of cosmic birefringence for the underlying theory and its consequences for the string swampland conjectures. We finally discuss oscillatory features in the dark energy sector and the expected cosmic birefringence tomography.  
}

\arxivnumber{}

\usepackage{amsfonts}

\begin{document}

\maketitle

\section{Introduction}
Identifying the microscopical origin of dark energy (DE) is one of the greatest challenges in modern physics \cite{CCPWeinberg}. One plausible explanation is the existence of an ultralight scalar field whose energy density is dominated by its potential, thus the equation of state is $\omega=p/\rho \ \simeq -1$ \cite{Peebles:2002gy,Copeland:2006wr,Tsujikawa:2010sc}. It's well known that axion-like particles\footnote{In this work, we refer to the ``axion-like particles" or ``axion field" almost exclusively as the dark energy field (namely, no fixed relationship between its mass and couplings), and hence distinguish it from the originally proposed ``QCD axion" \cite{Peccei:1977hh,Weinberg:1977ma,Wilczek:1977pj}.} with masses smaller than the current Hubble scale can play this role \cite{Frieman_1995, Kolda_1999,Svreck2006}. They are ubiquitous in string theory  \cite{2006JHEP...06..051S, Axiverse} and, being pseudo-Nambu-Goldstone-Bosons, can receive an arbitrary small mass from non-perturbative physics whose value is protected by the remnant shift symmetry at all orders in perturbation theory. Thus, axion-like particles could naturally explain DE, see e.g. \cite{Kim1999,Kim_2003,Choi2000,Kim2009,Marsh:2015xka}.

Axion-like particles can interact in a parity-violating manner with photons via the Chern-Simons coupling \cite{Sikivie:1983ip,Raffelt:1987im}. Hence, 
searching for parity-violating signatures in cosmological observations may help us to identify the nature of axions as DE. In the presence of axion-photon coupling, the displacement of the axion field along the light path induces a small difference in the phase velocity between the left- and right-handed photon that might contribute to what is known as  {\it cosmic birefringence} (CB) \cite{Carroll:1989vb,Carroll:1991zs,Harari:1992ea,Carroll:1998zi}. This effect rotates the polarization plane of linearly polarized light by an amount $\beta$, called birefringence angle (see \cite{komatsu2022new} for a recent review on this topic). In the standard cosmology its value is zero because it's usually assumed that our universe doesn't have a preferred chirality.

Cosmic microwave background (CMB) photons are an ideal target to probe CB because its potential effect accumulates over the cosmological distance from the Last Scattering Surface (LSS) \cite{Lue:1998mq}. Despite of many efforts to measure  $\beta$, however, its value could not be so accurately determined because of its degeneracy with a potential miscalibration angle of the polarimeters \cite{QUaD:2008ado,Komatsu_2011}.
Recently, the authors of \cite{minami2019simultaneous} have proposed a new technique to overcome this difficulty by using the polarization of the Milky Way foreground. With this method, they found a weak signal of isotropic birefringence angle $\beta = 0.35 \pm 0.14 \ \text{deg} \ (68\% \text{C.L.})$ from {\it Planck} public data release 3 (PR3) \cite{Minami:2020odp}. Cosmological implications of this result have been discussed in \cite{Fujita_2021,FujitaTOP,Takahashi_2021,Nagata:2021pvc,Fung:2021wbz,Mehta:2021pwf,CBDM2021,Jain_2021,Clark:2021kze,Fung:2021fcj,Namikawa:2021gbr,Alvey2021,Choi:2021aze,obata2021implications,sherwin2021cosmic,Mohammadi:2021xoh}.
An improved measurement of $\beta=0.30 \pm 0.11$ deg ($68\%$ C.L.) from PR4 has now been reported \cite{diegopalazuelos2022cosmic} which has lower noise and better-characterized systematics of the instruments. Accounting for the foreground EB correlation, the authors found $\beta=0.36\pm 0.11$~deg (68\%~C.L.), which exceeds the statistical significance of $3\sigma$. 

The accumulating evidences in favor of a non-zero CB represents a new and powerful tool to test DE models and to study the implications between observations and fundamental physics \cite{Fujita_2021, FujitaTOP, Alvey2021, Pogosian:2019jbt}.
In this paper, we consider a model of axion dark energy with the monodromy potentials motivated by string theory. Axion fields with monodromy potentials have been studied a lot in the context of inflation \cite{McAllister2008, Silverstein,Kaloper:2008fb,Flauger,Kaloper:2011jz,Kaloper:2014zba,Kaloper:2016fbr,DAmico:2021vka}, and in the case of late DE \cite{Kaloper:2008qs,Panda:2010uq,DAmico:2016jbm,DAmico:2018mnx}\footnote{In this note, we just consider a monodromy axion whose potential dominates the DE density. We do not address the notorious problem of explaining the smallness of the cosmological constant and we assume that the energy density at the minimum vanishes. However, axion monodromies have also been discussed in this context in \cite{Kaloper:2018kma}.}. In this scenario, the potential is monomial, thus the effective range of the axion field is enhanced compared to the standard periodic potential. This key feature allows to meet the slow-roll conditions for DE only with a super-Planckian field value, independently on the value of the decay constant \cite{Kaloper:2008qs,Panda:2010uq,DAmico:2016jbm,DAmico:2018mnx}. In the first model proposed for such axion DE the potential is linear \cite{Panda:2010uq}, thus we first analyze the implications of the CB measurement for such model in terms of the potential's slope. Subsequently, we extend the analysis to the more general class of monomial potentials and we also consider the effect of periodic modulations from subdominant instanton contributions. Remarkably, we find that the birefringence measurement provides a direct link between the slow-roll parameter and the axion decay constant that turns to be sub-Planckian with a maximum value around the Grand Unified Theory (GUT) scale $10^{16}$ GeV.  Therefore, we confirm that the axion monodromy can successfully explain the observed DE and birefringence angle and, simultaneously, it can relax the theoretical issues of the standard axion DE model such as the requirement of a super-Planckian decay constant \cite{Kaloper_2006} or the exponential fine-tuning of its initial conditions \cite{Hilltop2008}.
We further comment on a couple of new predictions whose detection would strongly support this kind of scenario.  

The paper is organized as follows.
In section \ref{sec:CBI}, we review the CB effect and its implications for the standard model of axion-like particles in section \ref{sec3}.
In section \ref{sec:monodromy}, we present the model of axion dark energy with monodromy potential and its implications for $\beta$.
In section \ref{sec:discussion}, we discuss our results in light of the string swampland conjectures, oscillations in the DE sector and the expectation for birefringence tomography \cite{sherwin2021cosmic}.
Finally, we summarize our work in section \ref{conc}. Throughout this paper, we set the natural unit $\hbar = c = 1$.

\section{Cosmic Birefringence of CMB Photons from Axion Dynamics} \label{sec:CBI}

In the presence of an axion field that interacts with photons through the Chern-Simons interaction, the axion-photon lagrangian reads:
\begin{equation}
\mathcal{L} = -\frac{1}{2}\partial_\mu \phi \partial^\mu \phi -\frac{1}{4} F_{\mu\nu}F^{\mu\nu} -\frac{1}{4}g_{\phi\gamma}\phi F_{\mu\nu}\Tilde{F}^{\mu\nu}- V(\phi),
\end{equation}
where $F_{\mu\nu}\equiv \partial_\mu A_\nu -\partial_\nu A_\mu $ is the field strength of photon and $\Tilde{F}^{\mu\nu}= \frac{1}{2}\epsilon^{\mu\nu\rho\sigma}F_{\rho\sigma}$ is its dual. The Chern-Simons coupling, $g_{\phi\gamma}\phi\Vec{E}\cdot\Vec{B}$, changes its sign under parity transformation since the electric and the magnetic fields have opposite parity properties.
Thus, this term affects the propagation of light in a parity-violating way.

We solve the dynamics in a spatially-flat Friedmann-Lema\^{i}tre-Robertson-Walker metric $\mathrm{d}s^2=-\mathrm{d}t^2+ a(t)^2\mathrm{d}\Vec{x}^2=a(\eta)^2[-\mathrm{d}\eta^2+ \mathrm{d}\Vec{x}^2]$.
In Fourier space, the equation of motion (EOM) for the circularly-polarized photon $A_\pm=(A_1 \mp iA_2)/\sqrt{2}$ is given by:
\begin{equation}\label{eq:EOMphtoton}
    (\partial_\eta^2+ \omega_{\pm}^2)A_{\pm}(\eta, k)=0, \qquad \omega_{\pm}^2 \equiv k^2\Big(1 \mp \frac{g_{\phi\gamma}}{k}\frac{\text{d}\phi}{\text{d}\eta}\Big) \ .
\end{equation}
The interaction with the axion field differentiates the dispersion relation between the two helicities of a polarized photon that leads to a different phase velocity. Indeed, after taking the square root of the second term in \eqref{eq:EOMphtoton}, we get $\omega_\pm \simeq k \mp (g_{\phi\gamma}\text{d}\phi/\text{d}\eta)/2$.
This equation has been computed under the assumption that $g_{\phi\gamma}\text{d}\phi/\text{d}\eta$ is much smaller than the spatial frequency of a wave $k$. So, at first order, the correction of the phase velocity is frequency independent \footnote{Note that higher orders in the expansion of the phase velocity do depend on the frequency, but Planck's sensitivity is far from detecting any frequency dependence on $\beta$ from the axion field because it enters at order $(g_{\phi\gamma}\text{d}\phi/\text{d}\eta)^3/k^2$ \cite{eskilt2022}. The author of this paper found no evidence for such dependence, that can be generated by Faraday rotation or by other beyond standard model theories, in the CMB data motivating even more to look for an explanation in this axion scenario.}.
As a consequence, the polarization plane of a linearly-polarized light, that can be equally decomposed into the two circularly-polarized modes, gets rotated during its propagation \cite{Carroll:1989vb,Carroll:1991zs,Harari:1992ea}. Assuming $\beta>0$ for a clockwise rotation, the total birefringence angle is given by: 
\begin{equation}\label{eq:Birefringence angle}
    \beta(\hat{n}) =\frac{1}{2} \int^{\eta_{obs}}_{\eta_{em}}\mathrm{d}\eta(\omega_- - \omega_+) = \frac{g_{\phi\gamma}}{2} \int^{\eta_{obs}}_{\eta_{em}}\mathrm{d}\eta \frac{\mathrm{d}\phi}{\mathrm{d}\eta} = \frac{g_{\phi\gamma}}{2}(\phi_{obs}( \hat{n})-\phi_{em}(\hat{n})) \ .
\end{equation}
Therefore, $\beta$ measures the difference in the field value between the end points of the light path and it is measurable once the polarization at these two moments is known. In this paper, we consider the isotropic birefringence because it is a good approximation for dark energy and because there is no evidence so far for the spatially varying one \cite{Namikawa:2020ffr,Bianchini_2020} \footnote{Future telescopes as SO, CMB-S4 and LiteBIRD have a better sensitivity thus can improve the bounds on} the anisotropic birefringence angle \cite{Pogosian:2019jbt,Greco:2022ufo,Zhai:2020vob}. The generations of anisotropic birefringence angle has been discussed in the literature on the constraints of isocurvature fluctuations \cite{Caldwell:2011pu,Fujita_2021}, early DE \cite{Capparelli:2019rtn}, or topological defects \cite{Agrawal_2020,Takahashi_2021,Jain_2021}.. 

Current measurements \cite{Minami:2020odp, diegopalazuelos2022cosmic,eskilt2022} look for CB searching for parity-violating signatures in the CMB polarization map. For instance, the rotation of the polarization plane of photons coming from LSS induces a mixing between the two polarization modes {\it E} and {\it B}, that have opposite parity properties. In the standard scenario, in which the statistical polarization distribution of CMB is isotropic, the parity-odd correlation $C^{EB}_l$ goes to zero after taking the sky average \cite{Zaldarriaga_1997, Kamionkowski_1997}. Thereafter CB induces a rotation of the polarization plane which leads to an observed $C^{EB}_l$, that, in the case of isotropic rotation, is related to the original power spectra via $\beta$ in the following way \cite{Lue:1998mq,Feng_2005,Feng:2006dp,Liu:2006uh,Finelli_2009,Lee:2014rpa} \footnote{This formula assumes $C^{EB,CMB}=0$ as in the standard case, but it should be introduced if there were parity-violating phenomena occurring in the early Universe such as chiral gravitational waves \cite{Lue:1998mq,Saito:2007kt,Contaldi:2008yz,Sorbo2011,Namba:2015gja,Thorne:2017jft,Li_2008}.}:   
\begin{equation}
    C^{EB,obs}_l=\frac{\sin(4\beta)}{2}(C_l^{EE,CMB}-C_l^{BB,CMB}) \ ,
\end{equation}
where we have denoted ``$CMB$" as the intrinsic spectra at the time of LSS. Therefore, from the analysis of the observed {\it EB} and the theoretically-known $EE$ and $BB$ power spectrum, $\beta$ is inferred. Currently, the greatest challenges for the measurement are the lack of knowledge on the potential miscalibrations of the polarimeters \cite{Minami:2020odp, minami2019simultaneous} and the uncertainties in modeling the {\it EB} foreground of the Milky Way \cite{Clark:2021kze}. In particular, this last one is the reason for which no cosmological significance to the latest result has been assigned \cite{diegopalazuelos2022cosmic}.

\section{Axion Phenomenology from Isotropic Cosmic Birefringence}\label{sec3}
It follows from \eqref{eq:Birefringence angle} that in the case of a homogeneous axion field the birefringence angle is isotropic and is given by the evolution of the background field from LSS until now:
\begin{equation}\label{eq:betaformula}
\beta=\dfrac{g_{\phi\gamma}}{2}\Delta\phi \ , \qquad \Delta\phi \equiv \phi_{0}-\langle \phi_{\text{LSS}}\rangle \ ,
\end{equation}
where the last term is the field average over the finite thickness of recombination weighted by the visibility function \cite{Fedderke:2019ajk,Fujita_2021}.
From the measured value of $\beta$, we can infer the expected value of the coupling constant $g_{\phi\gamma}$ by computing the field displacement $\Delta\phi$ for different values of the model parameters. Thus, we need to solve the EOM for the background field coupled with the Friedmann equations:
\begin{align}\label{eq:EOMscalar}
    &\ddot{\phi} + 3H\dot{\phi} + \dfrac{\text{d}V}{\text{d}\phi} = 0 \\
    &H = H_0\sqrt{\Omega_r(1+z)^4 + \Omega_m(1+z)^3 + \Omega_\Lambda + \Omega_\phi} \ ,
\end{align}
where the dot indicates the time derivative and $\{\Omega_r, \ \Omega_m, \ \Omega_\phi, \ \Omega_\Lambda\}$ are the density parameters of radiation, matter, cosmological constant, and the axion component. The axion abundance is usually chosen to saturate the respective bounds in the different mass regions coming from the observations of CMB and large scale structure \cite{Hlozek:2014lca}:
\begin{equation}
\Omega_{\phi,max}= 
\biggl\{
\begin{aligned}
&0.69 & m< 8.5 \times 10^{-34} \text{eV}\\
&0.006h^{-2} & 10^{-32}\text{eV} < m < 10^{-25.5} \text{eV}.
\end{aligned}
\end{equation}
where $h=0.677$. In the first line the axion abundance plays the role of the whole DE (thus we take $\Omega_\Lambda=0$) whereas in the second line it represents just a tiny fraction of dark matter. In the intermediate region the value of $\Omega_{\phi,max}$ can be simply interpolated between the two maximum values. The maximum axion mass considered in this equation is the maximum value consistent with CB observation as we will explain below.
It is customary to consider the axion field evolving in a cosine potential written in terms of the mass $m_a$ and the decay constant $f_a$:  
\begin{equation}\label{eq:cosinepot}
V(\phi) = m_a^2f_a^2\left[1-\cos\left(\dfrac{\phi}{f_a}\right)\right] \ ,
\end{equation}
which reduces to the quadratic form $V \simeq m_a^2\phi^2/2$ in the small field range. With this potential, there are three mass regimes in which the field displacement, namely the inferred coupling, has a different dependence on the parameter values $\{m_a,\Omega_\phi\}$ \cite{FujitaTOP}. This is determined by the onset of the field oscillations occurring at $H_{osc}\sim m_{osc}$ relative to the LSS $H_{\text{LSS}}\sim 10^{-29}$ eV.
For $m_a \lesssim H_0\sim 10^{-33}$ eV, the field displacement decreases inversely proportional with the mass, thus the inferred value of $g_{\phi\gamma}$ increases until exceeding the current allowed value. Using the Chandra bound of $g_{\phi\gamma}<1.4\times 10^{-12}\text{GeV}^{-1}$ \cite{Berg:2016ese}, a lower limit for the axion mass can be inferred $m_a \sim 10^{-41}$ eV \cite{FujitaTOP}. The range of masses $H_0 \lesssim m_a \lesssim H_{\text{LSS}}$ is the typical one addressed for CB because it leads to the greatest $\Delta\phi$. This happens because the field starts oscillating around the minimum and its amplitude scales as $a^{-3/2}$ between the LSS and today, which makes $\phi_0$ negligible compared to $\phi_{\text{LSS}}$, thus $\Delta\phi\simeq -\langle \phi_{\text{LSS}}\rangle$. For greater masses $m_a\gtrsim H_{\text{LSS}}$, the rapid oscillation of the field during recombination exponentially suppresses the averaged value of $\langle \phi_{\text{LSS}}\rangle$\footnote{
See \cite{Fedderke:2019ajk} for the phenomenology birefringence in the presence of oscillating axion field.}. Eventually, for $m_a\gtrsim 2.7 \times 10^{-27}$ eV, $\phi_0$ becomes dominant and $\Delta\phi$ gets proportional to the inverse of $m_a$. The corresponding axion-photon coupling is proportional to  $m_a$ and it exceeds the Chandra constraint for masses greater than $m_a\gtrsim 10^{-25}$ eV. To conclude, the observation of CB can be explained by axion with two mass regimes: $m_a \in (10^{-32},10^{-25})$ eV, axions as a tiny fraction of dark matter\footnote{For instance \cite{CBDM2021} discussed a scenario in which axions acquire naturally this typical range of masses when they are coupled with dark matter energy density. Authors of \cite{Mehta:2021pwf} have also argued that axions in this typical range naturally emerge from a vast set of compactifications.
Early DE axions can also explain CB as shown in \cite{FujitaTOP} and their dynamics could ease the Hubble tension as opposed to the late DE evolution \cite{Banerjee:2020xcn,Heisenberg:2022gqk,Heisenberg:2022lob}.}, and $m_a \in (10^{-41},10^{-33})$ eV, axions as DE \cite{FujitaTOP}.

Despite the success in explaining $\beta$, axion DE with standard cosine potential suffers from some theoretical issues. For the quadratic potential, the requirement of being in slow-roll regime demands a super-Planckian field value: 
\begin{equation}
    \frac{\phi_0}{M_{Pl}}\simeq \sqrt{6\Omega_\phi}\frac{H_0}{m_a}\gg 1 \qquad \longleftrightarrow \qquad m_a\ll H_0 \ ,
\end{equation}
that  is  inconsistent  with  the  quadratic  potential  being  the  approximation  of  the  full  non-perturbative cosine potential \eqref{eq:cosinepot}. In the last case instead, the slow-roll condition demands a super-Planckian decay constant \cite{Kaloper_2006}:
\begin{equation}
\dfrac{\partial_\phi^2V}{V}M_{Pl}^2 \ll 1 \qquad \longleftrightarrow \qquad f_a \gg M_{Pl} \ .
\end{equation}
This is not favored in string theory since the higher harmonics of instantons with order of $f_a/M_{Pl}$ would spoil the flatness of the potential \cite{Banks:2003sx}. Moreover, strong arguments, as gravity as the weakest force \cite{Arkani_Hamed_2007}, prefer values of the decay constant around two orders of magnitude smaller than the Planck scale $f_a \lesssim {M_{Pl}}/{S_{ins}}\sim 10^{-2}M_{Pl}$, where $S_{ins}$ is the instanton action \cite{Ibe_2019}. In this case, explaining the observed DE behaviour through an axion field demands an exponential fine tuning of its initial value close to the top of the potential \cite{Hilltop2008}. The latter results very unnatural in the context of axion misalignment production and the isocurvature fluctuations \cite{Kaloper_2006,Cicoli:2021skd}\footnote{Recent attempts to explain this maximal-misalignment mechanism dynamically \cite{stochastic2021} or to achieve an effective $f_a \sim \mathcal{O}(1)M_{Pl}$, that highly relaxes the above fine-tuning issues \cite{Kaloper_2006,Svreck2006,Ibe_2019,Kamionkowski_2014,Kim1999,Kim_2003}, have been explored. Another possibility to achieve a super-Planckian decay constant relies on the multi-field scenario, indeed in \cite{obata2021implications} the case in which a combination of two fields makes the dark energy and the other the dark matter field content has been discussed in connection with the various birefringence experiments.}. In the next sections we explain how the above issues can be avoided by axion models with monodromy potentials.

\section{Implications for Monodromic Axion Dark Energy}
\label{sec:monodromy}

In this section, we first study the implications from CB for the original model of DE with linear potential by Panda, Sumitomo and Trivedi \cite{Panda:2010uq}. Here the central parameter is just the slope of the potential that we will denote with $s$.  Subsequently, motivated by the previous studies in the context of axion monodromy inflation  \cite{McAllister2008, Silverstein,Flauger,Kaloper:2008fb,Kaloper:2011jz,Kaloper:2014zba,Kaloper:2016fbr,DAmico:2021vka} and late DE \cite{Panda:2010uq,Kaloper:2008qs,DAmico:2016jbm,DAmico:2018mnx}, we consider the broader class of  monomial potentials \cite{monodromypower}. We found that these models, in the slow-roll limit, can be commonly described just by the present value of the axion field that must be super-Planckian.

\subsection{Case A: linear potential}
The authors of \cite{Panda:2010uq} have shown that axion monodromy gives a workable model for axion DE with a super-Planckian field value and a sub-Planckian decay constant. The fundamental idea is that the axion shift symmetry is mainly broken by the interplay of branes and axions in the same region of the internal space.
The model set-up consists of an $NS5$ brane and an $\overline{NS5}$ anti-brane positioned in two highly warped throats in the internal space where the axion, the zero mode of RR 2-form field $C_2$, is defined.
The resultant potential of this configuration, coming from the Dirac-Born-Infeld action of the branes in the presence of the antisymmetric 2-form $C_2$, gives rise to a potential for the axion that is linear in the large field limit. Moreover, they argue that the linear behaviour is not spoiled by higher corrections coming from the embedding of the model in the complete UV framework.  The potential can be parametrized by: 
\begin{equation}
    V= \mu^4\frac{\phi}{f_a} \ ,
\end{equation}
where the energy scale $\mu$ is determined by the warped factor at the bottom of the throat that can be successfully adjusted to match the present DE density $\mu \sim 10^{-3}\text{eV}$.
More detailed discussion in terms of string parameters can be found in the original paper \cite{Panda:2010uq}.

In this setup, the central parameter that controls the cosmological evolution of the background field is just the slope of the potential that we define as:
\begin{equation}
s=\frac{1}{3M_{Pl}^2H_0^2}\frac{\mathrm{d}V}{\mathrm{d}\phi}=\frac{\mu^4/f_a}{3M_{Pl}H_0^2} \ ,
\end{equation}
such that the axion abundance is approximately linear in the axion field $\Omega_\phi\simeq s\phi$. As before, we now compute the field displacement $\Delta\phi$ for different values of this parameter to infer $g_{\phi\gamma}$ from the measurement of $\beta$\footnote{The evolution of the axion field and its success in fitting the cosmological observables have been discussed also in \cite{Gupta2012}, but here we want to find a direct relation between the field displacement to the parameter of the potential. }. 

The evolution of the axion field from LSS is given by solving the corresponding EOM coupled to the Friedman equation \eqref{eq:EOMscalar}:
\begin{equation}\label{eq:MONOeom}
    \phi_n''+ 3\mathcal{H}\phi_n'+3s=0 \ ,
\end{equation}
where the derivative is with respect to the dimensionless time variable $\tau = H_0 t $ and we have defined $\mathcal{H}=H/H_0$ and $\phi_n=\phi/M_{Pl}$. We take the initial value of the field $\phi_i$ such that the final axion abundance matches with the one of DE $\Omega_\phi=0.69$ in a flat universe \cite{Planck:2018vyg}. Thereafter, we consider that the axion field explains the totality of the observed DE density, thus we assume a vanishing vacuum energy $\Omega_\Lambda=0$. Moreover, we set the initial velocity as $\dot{\phi}_i=0$ because the big drag of the Hubble term freezes the field at its initial value independently on its initial velocity.

\begin{figure}
    \centering
    \begin{minipage}[b]{0.45\linewidth}
    \includegraphics[width=6.8cm]{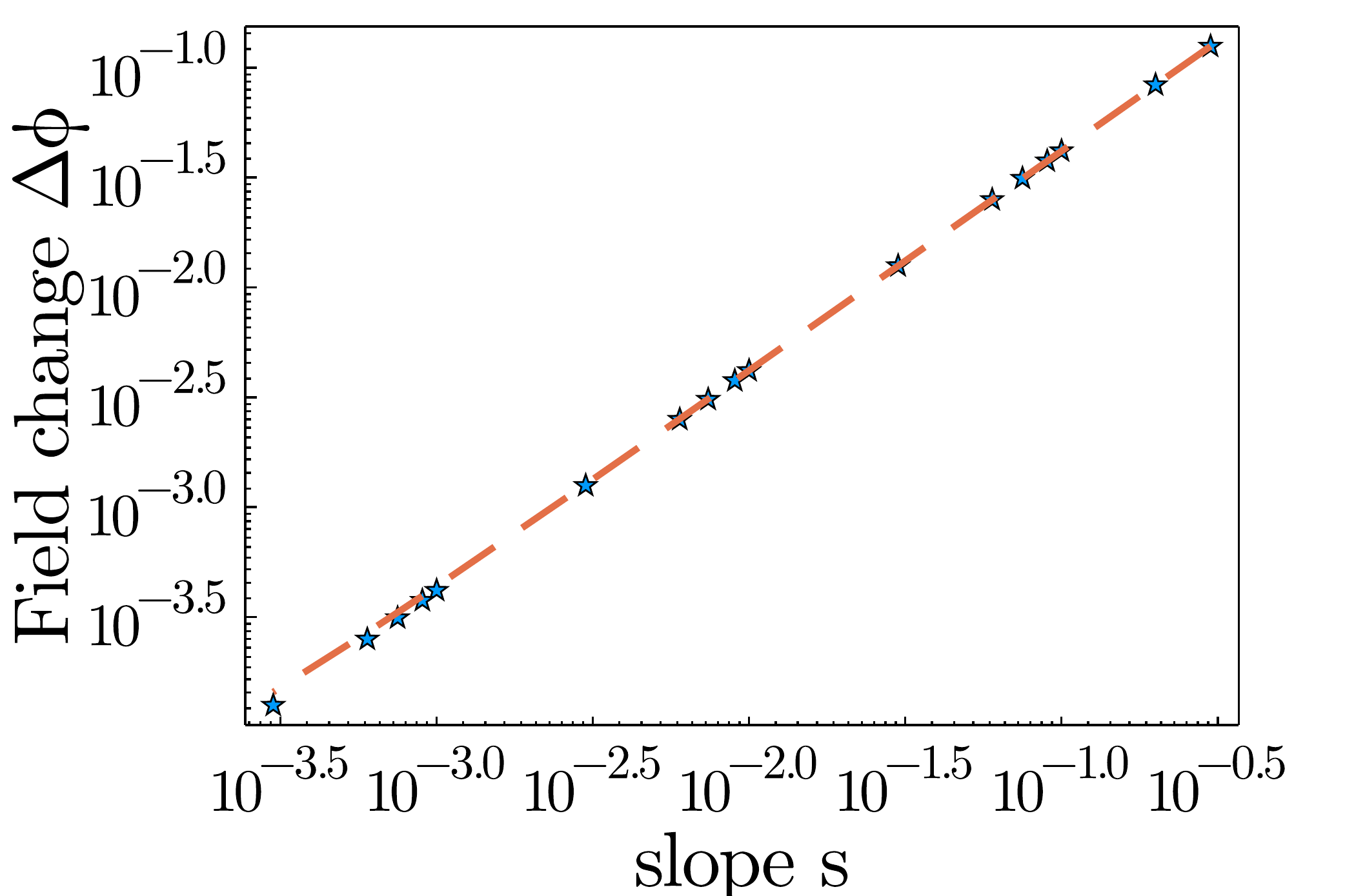}
    \end{minipage}
    \quad
    \begin{minipage}[b]{0.45\linewidth}
    \includegraphics[width=6.8cm]{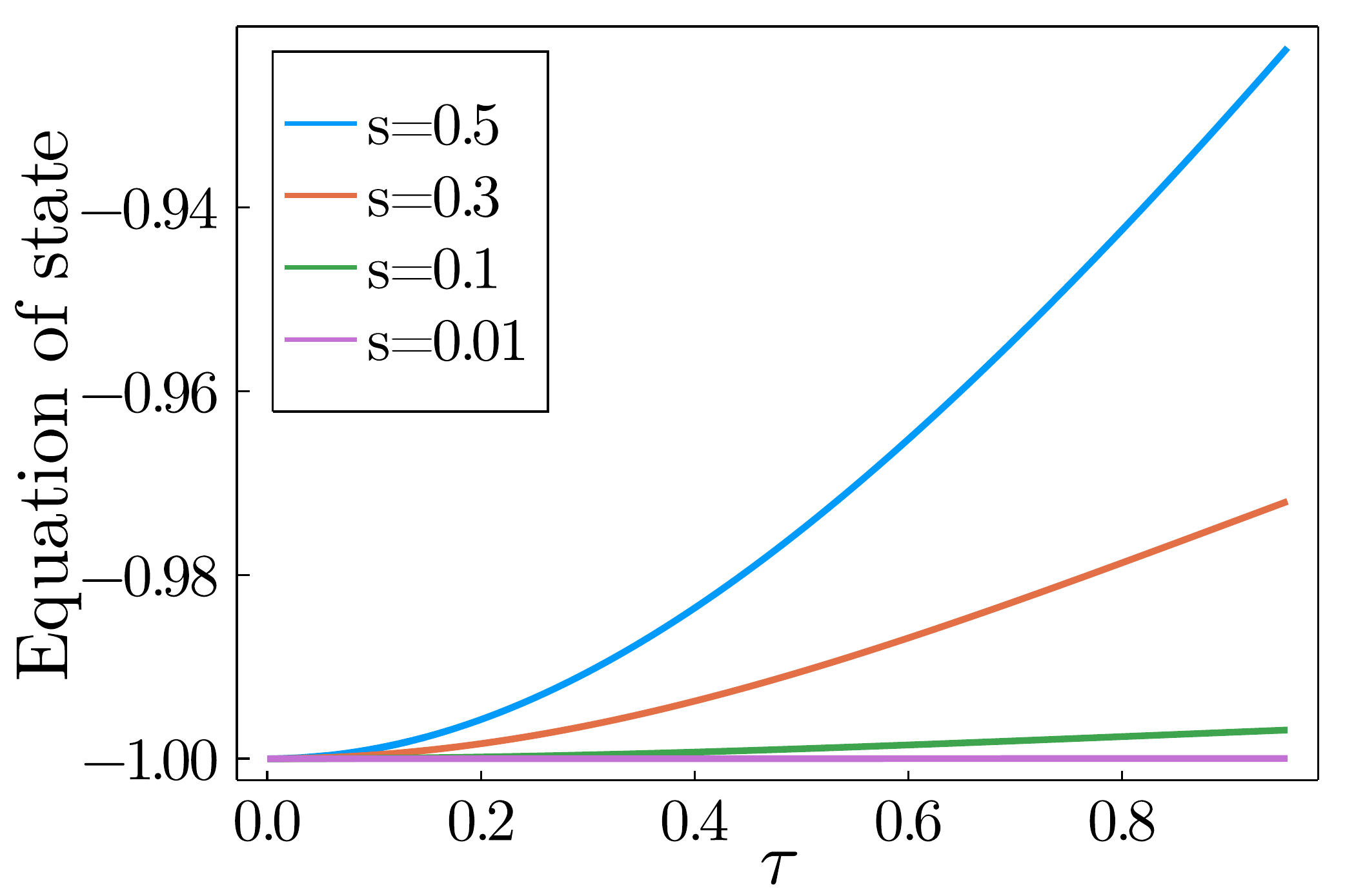}
    \end{minipage}
    \caption{On the left panel we show the linear relation between the absolute value of the field change and the slope of the potential. The right panel displays the evolution of the equation of state as a function of the dimensionless time coordinate $\tau=H_0t$ for the sample values of $s= \{0.5,0.3, 0.1, 0.01\}$. The corresponding initial values are chosen to match the current DE abundance and are $\phi_{i,n}=\{1.54, 2.4, 6.94, 69\}$. The evolution of the field is given by the numerical solution of the EOM \eqref{eq:MONOeom} with values $\Omega_m=0.31$, $\Omega_\phi= 0.69$ and $a_{eq}=1/3400$.}
    \label{fig:evolution abundance and EOS}
\end{figure}

\begin{figure}
    \centering
    \includegraphics[scale=0.6]{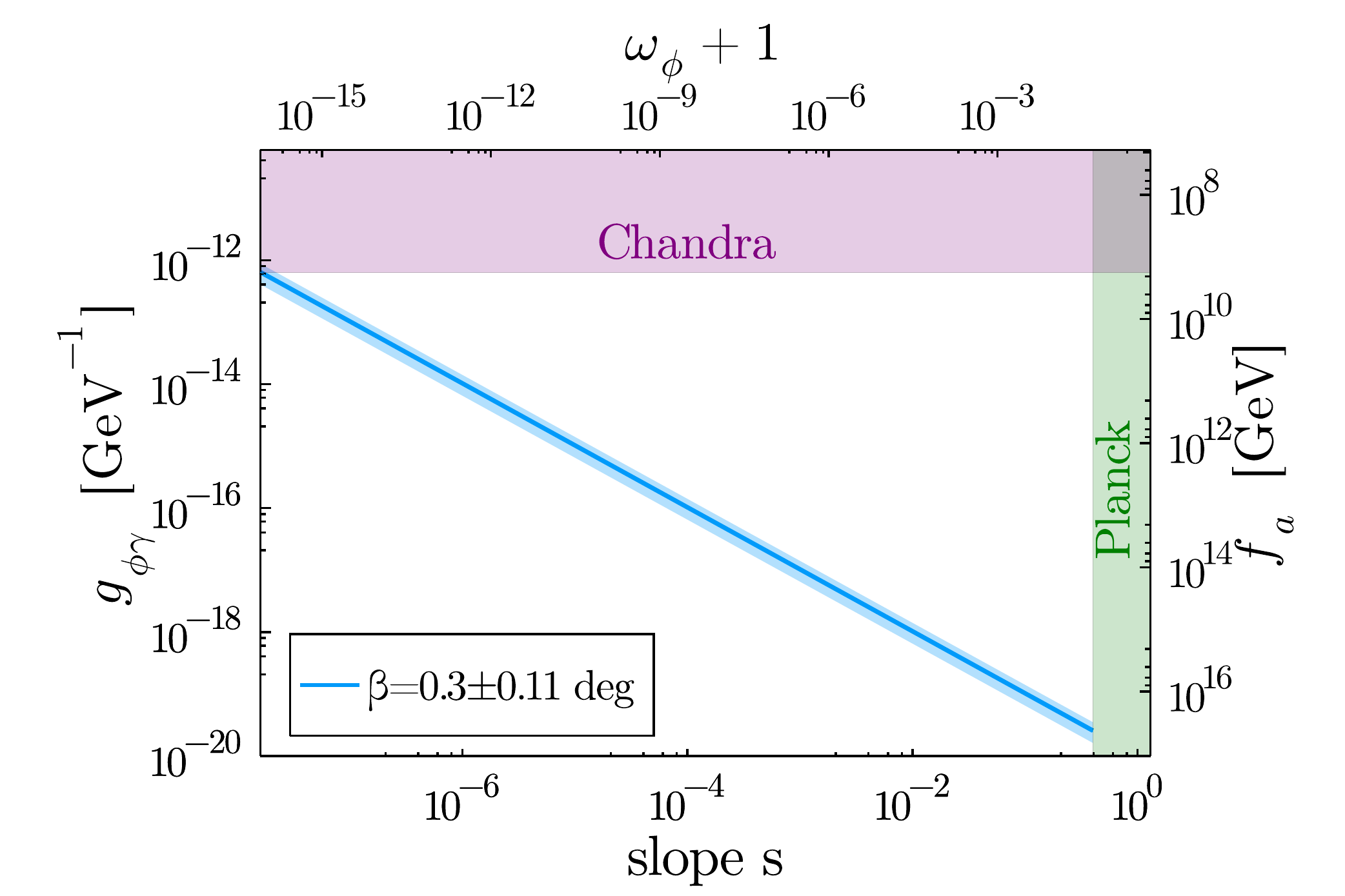}
    \caption{The axion-photon coupling constant inferred from the linear potential $V/\rho_c= s\phi_n$ as a function of the slope of the potential (bottom x-axis) and the final equation of state (top x-axis). The inferred axion decay constant is also displayed (right y-axis) as a function of these quantities. Smaller slopes are ruled out by the current constraint $g_{\phi\gamma}<6.3 \times 10^{-13}\text{GeV}^{-1}$ \cite{reynes2021new} whereas bigger values are at odds with Planck measurements $\omega_\phi+1 \leq 0.05$ \cite{Planck:2018vyg}. The light blue area shows the uncertainty on the birefringence angle $\beta= 0.30\pm0.11$ deg  \cite{diegopalazuelos2022cosmic}. } 
    \label{fig:coup}
\end{figure}

For the results showing below we solve numerically  \eqref{eq:MONOeom}, but we first discuss the analytical solution of the EOM in matter-dominated period since the numerical solution starts deviating from it only recently when the DE component becomes dominant. Indeed, because the axion field evolves mainly in matter-dominated period, the numerical results give just small corrections to the equations presented below. Substituting $\mathcal{H}= 2/(3\tau)$ in eq. \eqref{eq:MONOeom}, it's easy to check that the correct solution is given by $\phi_n'=-s\tau$ \cite{Weinbergcos}. Thus the field evolution and its displacement from LSS is given by:
\begin{equation}\label{eq:MDsolution}
\phi_n(\tau) = \phi_{i,n} - \dfrac{s\tau^2}{2} \quad \longrightarrow \quad \Delta \phi_n = -\dfrac{s}{2}(\tau_0^2 - \tau_{\text{LSS}}^2) \simeq  -\dfrac{s}{2}\tau_0^2 \ .
\end{equation}
The age of the universe depends on the various cosmological parameters, but for the case considered here its value is very close to that of $\Lambda$CDM, $\tau_0=0.95$ \footnote{Here we consider $\Omega_m=0.31$, $\Omega_\Lambda= 0.69$ and $a_{eq}=1/3400$.}. The linear dependence between $\Delta\phi$ and the $s$-parameter is shown in the left panel of figure \ref{fig:evolution abundance and EOS}. From the numerical interpolation, we find the precise coefficient of proportionality: $|\Delta \phi_n|= 0.417s$. In what follows, we always refer to the absolute value of $\Delta \phi_n$ for simplicity, but we keep in mind that it's actually negative because the field value decreases in time. Note that the linear relation \eqref{eq:MDsolution} doesn't depend on the axion abundance or on the initial conditions, meaning that $\beta$ truly measures the local steepness of the potential encoded by the $s$-parameter. It follows that our result can be regarded as a first approximation for a generic nearly-flat potential around a given value $\phi_\star$, where $s$ is defined via:
\begin{equation}
    \Omega_\phi \simeq s(\phi-\phi_\star)+c \quad \text{with} \quad s=\frac{1}{\rho_c}\frac{\mathrm{d}V}{\mathrm{d}\phi}\Bigr|_{\phi_\star} \,
\end{equation}
where $c=\Omega_\phi(\phi_*)$ and $\rho_c=3M_{Pl}^2 H_0^2 $ is the critical energy density.
The DE equation of state also depends on the local steepness of the potential as shown in the right panel of \ref{fig:evolution abundance and EOS}. In particular, the equation of state is frozen at $\omega_\phi=-1$ at early times because of the Hubble drag and then it starts raising as:
\begin{equation}
\omega_\phi+1 \simeq \dfrac{\dot{\phi}^2}{V} \simeq \dfrac{s^2\tau_0^2}{3\Omega_\phi} \ , \label{eq: omegas}    
\end{equation}
where the solution in matter-domination is used. Numerically, we find that the deviation from cosmological constant behaviour is related to the square of $s$ via $\omega_\phi+1= 0.31s^2$.\footnote{The numerical result is lower than the analytic one because the solution in matter dominated era overestimates the evolution of $\omega_\phi$ at late times when the effective DE density becomes the dominant component.} Therefore, we can use the result from {\it Planck} 2018 data $\omega_\phi<-0.95$ at $95\%$ CL \footnote{This constraint comes from the best fit of the combined analysis with \textit{Planck}+lensing+SNe+BAO data, but hereafter we just refer to it as Planck bound.} \cite{Planck:2018vyg} to find an upper bound for the slope of the potential that is $s\leq0.4$. 
From the relation with the axion-photon coupling $g_{\phi\gamma}=2\beta/\Delta\phi$,  we obtain the following numerical result:
\begin{equation}\label{eq:coupres}
    g_{\phi\gamma}= 2.57\times
    10^{-20}\text{GeV}^{-1}\Big(\frac{|\beta|}{0.30\text{deg}}\Big)\Big(\frac{0.4}{s}\Big) 
\end{equation}
This can be equivalently expressed in terms of the final equation of state via \eqref{eq: omegas}:
\begin{equation}\label{eq:coupres1}
    g_{\phi\gamma}=2.57 \times 10^{-20}\text{GeV}^{-1}\Big(\frac{|\beta|}{0.30\text{deg}}\Big)\Big(\frac{0.05}{\omega_\phi+1}\Big)^{1/2}\Big(\frac{0.69}{\Omega_\phi}\Big)^{1/2}.
\end{equation}
The results are also shown in figure \ref{fig:coup}, together with the current excluded region of the axion-photon coupling.
To our best knowledge, the current tightest constraint comes from the study of the spectral distortion of the quasar $H1821+643$ from \textit{Chandra} observations \cite{reynes2021new}, from which the non-detection of the spectral distortion ascribable to axion-photon conversion demands the coupling smaller than $g_{\phi\gamma}<6.3 \times 10^{-13}\text{GeV}^{-1}$ at $99.7$\% for axions-like particles with masses $m_a<10^{-12}$ eV. 
This constraint translates into a lower bound for the steepness of the potential $s\geq 1.6\times10^{-8}$ and for the final equation of state:
\begin{equation}\label{eq:EOSbound}
    \omega_\phi+1\geq 2.67\times 10^{-16}\Big(\frac{|\beta|}{0.30\text{deg}}\Big)^2.
\end{equation}
This value is larger than what was found in \cite{FujitaTOP} because the  relationship between $\omega_\phi+1$ and $\Delta\phi$ is slightly different from what is found by slow-roll approximation and because we have used another updated bound \cite{reynes2021new}. Note that eq. \eqref{eq:coupres} depends only on the slope of the potential whereas its connection with the equation of state introduces the dependence on the axion abundance, as shown in eq. \eqref{eq:coupres1}
This happens because the equation of state is mainly sensitive to the current field value $\phi_0$ which determines the importance of the $\dot{\phi}/V$ term as we discuss shortly.

In addition, the relations \eqref{eq:coupres} and \eqref{eq:coupres1} give a new connection between the dark energy parameters and the axion decay constant $f_a$, for which we have not made any assumption so far.
Indeed, considering the ordinary relation of the coupling:
\begin{equation}\label{eq:coupling&fa}
    g_{\phi\gamma}= \frac{\alpha_{em}c_{\gamma \phi}}{2\pi f_a} \ ,
\end{equation}
in terms of the electromagnetic fine structure constant $\alpha_{em}$ and the anomaly coefficient $c_{\gamma \phi}$, we obtain: 
\begin{equation}
    \frac{f_a}{c_{\gamma \phi}}= 4.52 \times 10^{16}\text{GeV}\Big(\frac{0.30\text{deg}}{|\beta|}\Big)\Big(\frac{s}{0.4}\Big)  
\end{equation}
or, equivalently:
\begin{equation}\label{eq:decayres}
    \frac{f_a}{c_{\gamma \phi}}= 4.52  \times 10^{16}\text{GeV}\Big(\frac{0.30\text{deg}}{|\beta|}\Big)\Big(\frac{\omega_\phi+1}{0.05}\Big)^{\frac{1}{2}}\Big(\frac{0.69}{\Omega_\phi}\Big)^{1/2}.
\end{equation}
Remarkably, $\omega_\phi+1\leq 0.05$ implies an upper bound for the decay constant on the order of the GUT scale that, as mentioned before, it is in the preferable range of string theory computations. This is one of our main results, suggesting that the monodromic axion is a consistent single-field model of DE with a sub-Planckian decay constant that is simultaneously compatible with the cosmological measurements of the equation of state and CB. 

\subsection{Case B: general monomial potential}
Here, we show that these results are not specific of the linear model, but they can be generalized to a broader class of monomial potentials \cite{monodromypower}. Indeed, axion monodromy predicts a generic potential of the type:
\begin{equation}\label{eq:monomials}
    V=\mu^{4-n}\phi^n,
\end{equation}
with $n$ an integer or rational number \cite{monodromypower}. This class is known in the literature with the name of the \textit{thawing} class \cite{Caldwell2005} and it's characterized by a recent departure from $\omega_\phi=-1$. 

It is understood that, when the fields is deeply in the slow-roll regime $(\partial_\phi V/V)^2M^2_{Pl}\ll1$, all these models approach to a single generic behaviour uniquely characterized by the slow-roll parameter $\lambda_0=M_{Pl}\partial_\phi V/V=n/\phi_{n,0}$ and the  abundance $\Omega_\phi$ \cite{thawingquintessence}. As $\lambda_0$ approaches zero the overall evolution becomes almost indistinguishable for different DE models \cite{quintlaststand}. In particular, demanding $\omega_\phi+1\leq 0.05$ for the potential \eqref{eq:monomials} requires just a field value higher than the Planck scale $\phi_{n,0}/n\gtrsim 1$, independently on the value of $\mu$. This finds a natural explanation in the axion monodromy scenario.

Thus, we can repeat the exercise of the previous section and connect the field displacement needed to generate the CB observation to the important parameter determining the field evolution $\lambda_0=n/\phi_{n,0}$. The main behaviour of $\Delta\phi_n$ and $\omega_\phi+1$ follow from the naive slow-roll approximation \footnote{We point out that, even if the slow-roll approximation captures the correct dependence on the slow-roll parameter, it is not accurate for late DE because the field evolution occurs mainly in matter dominated period.
This makes less straightforward to relate the equation of state and the evolution of DE to the slow-roll potential parameters $\lambda_0$ \cite{SR2006}. } $\dot{\phi}^2/V\simeq \lambda_0^2\Omega_\phi/3$. We find that numerical interpolation gives the following accurate relations:
\begin{equation}\label{eq:omegaalpha}
    \omega_\phi+1 ={0.149}{\lambda_0^2} \hspace{2cm}  \Delta\phi_n={0.29}{\lambda_0}= 0.75\sqrt{\omega_\phi+1}.
\end{equation}
The first equation is close to the results in \cite{thawingquintessence} and the second relation is compatible with the results discussed in \cite{quintlaststand,PoleDE}.  The reason why these formulas are precise for different powers of the potential is that the field never rolls for long, thus it cannot see the full shape of the potential \footnote{ Moreover the characteristic relation between $(\omega_0, \omega_a)$ \cite{quintlaststand, Marsh2014} for this class of models could help in distinguish this from other scenarios.}. Note that the above formulas are consistent with the results of the linear potential via $s\simeq \lambda_0 \Omega_\phi$. Indeed, relating \eqref{eq:omegaalpha} to the observed $\beta$ gives again the same results as before \eqref{eq:coupres1} and \eqref{eq:decayres}. Therefore, the one-to-one correspondence between the equation of state and the $f_a$ holds for all this class of models.
Moreover, the condition $\omega_\phi+1\leq 0.05$ bounds the present field value to be $\phi_{0,n}\geq 1.7 n$, whereas constraint on the axion-photon coupling \cite{reynes2021new} gives the upper bound $\phi_{0,n}<2.3\times 10^{7}n$.
Such a very high value might not be favorable in the string theory framework, but nevertheless it represents interesting implications for the underlying UV theory. 

\section{Discussion}\label{sec:discussion}
\subsection{Bounds for the string swampland}
Note that eq. \eqref{eq:EOSbound} is currently the tightest lower bound on the deviation from cosmological constant behaviour of DE. This is particularly intriguing in light of the string swampland program that tries to extract the generic features of the effective field theories that are fully consistent with a quantum theory of gravity. In this regard, de Sitter space is very hard to construct and therefore a dynamical field seems a more natural explanation for the acceleration of the Universe \cite{Dvali:2014gua,Dvali:2017eba,Danielsson:2018ztv,Swampland2018}. 

String theory suggests two criteria between the field displacement and the slow-roll parameter for the corresponding scalar field not to be in the Swampland \cite{Obied:2018sgi,Garg:2018reu,Ooguri:2018wrx}.
The first criterion states that \textit{the traversed field range is bounded by  $\Delta\phi \sim \mathcal{O}(1)M_{Pl}$}.
As we can see from eq. \eqref{eq:omegaalpha} the considered models predict the field rolling over sub-Planckian distances and therefore naturally fulfill this criterion.
The second one asserts that \textit{there is a lower bound on $\partial_\phi V/V > c/M_{Pl} \sim \mathcal{O}(1)$ in any consistent theory of gravity when $V>0$} \cite{Swampland2018}. 
This means that the slope of the potential should be around the same order of the potential for a consistent model, but Planck data already bound $c\leq0.58$ to be smaller than one \footnote{Here we keep the notation of \cite{Swampland2018} in discussing the conjectures, but the $c$-parameter is directly related to the slow-roll parameter $\lambda_0$ defined previously.}.
On the other hand, CB provides the only way to infer a lower bound on $c$ from observations. We note that from eq. \eqref{eq:EOSbound} it follows a very small lower bound $c\simeq 4\times 10^{-8}$.
Surely, a deeper discussion in light of the full string theory model is needed for the feasibility of the corresponding high field values required for such small $c$-parameter.
Nevertheless, these results, together with the upper bound on the axion-decay constant \eqref{eq:decayres}, demonstrate the power of CB in probing string theory phenomenology for which other cosmological probes, as large scale structure and distances, are not sensitive enough. Improving the observational constraints for the axion-photon coupling and the equation of state will narrow further the current parameter space of these quantities.

\subsection{Oscillating dark energy}
Now we consider the consequence of adding, on top of the linear potential,  periodic contributions coming from the instanton effects. This scenario is not only phenomenologically interesting, in the case of inflation it leaves oscillations in the power spectrum, but it's proved to be generic for monodromies \cite{Flauger}. Following the previous literature, we write the potential as:
\begin{equation}\label{eq:modulatedpotential}
    V(\phi)= \frac{\mu^4}{f_a}\Big(\phi+b f_a \cos\Big(\frac{\phi}{f_a}\Big)\Big).
\end{equation}
We consider simply the case in which the potential is monotonic so that the modulations are small and the field would not get trapped in a local minimum, this implies $b<1$. In particular, we are interested in the case where the axion field rolls over several fundamental periods during its evolution because that could leave oscillations in the equation of state and in the DE abundance that are characteristic of the monodromy potential. We refer to \cite{schmidt2017monodromic} for the discussion on the observational signatures of such monodromic DE and we instead focus on the implications of a future detection for the axion decay constant and the anomaly coefficient. 

\begin{figure}
    \centering
    \begin{minipage}[b]{0.40\linewidth}
    \includegraphics[width=6.5cm]{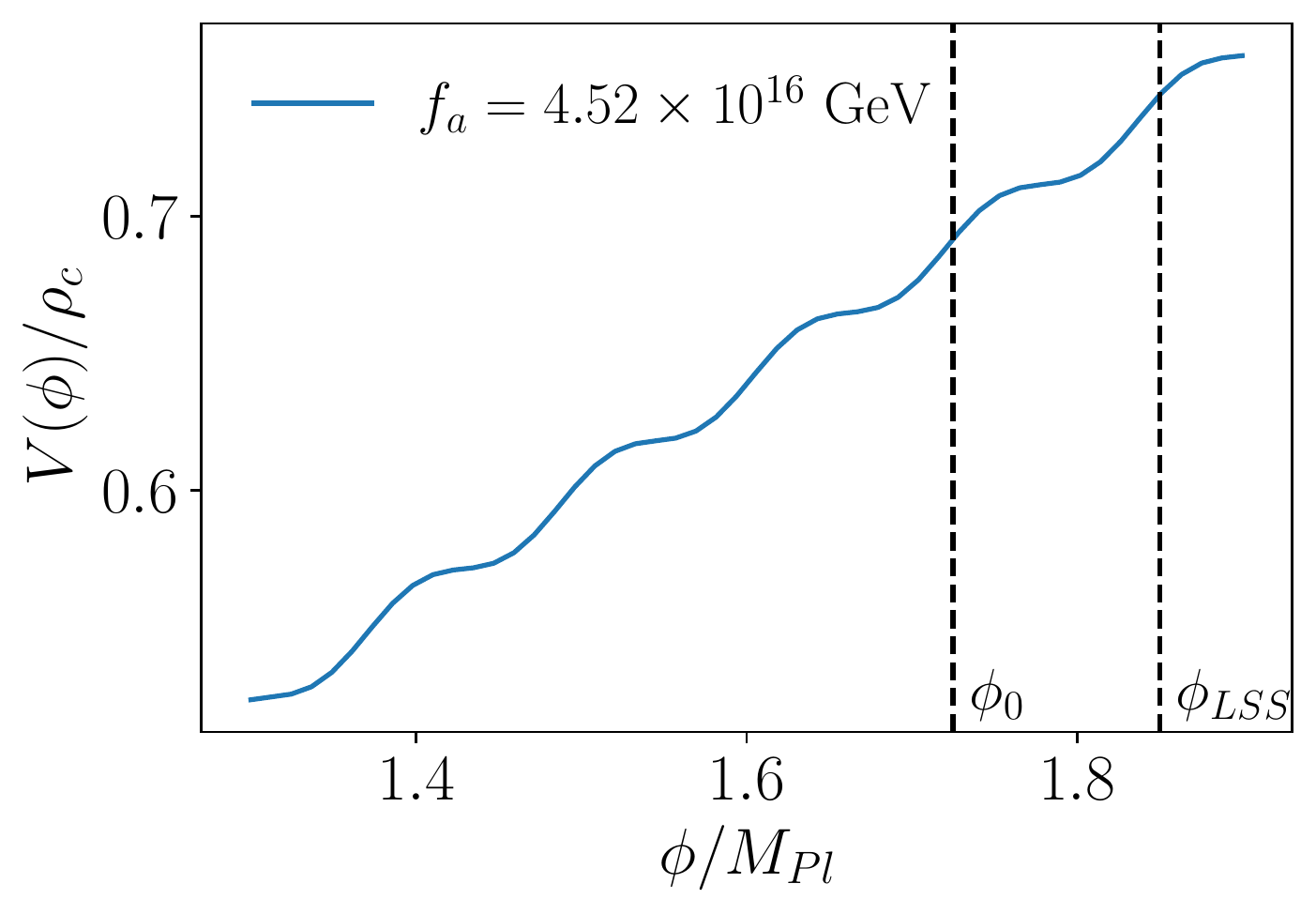}
    \end{minipage}
    \quad
    \begin{minipage}[b]{0.50\linewidth}
    \includegraphics[width=7.5cm]{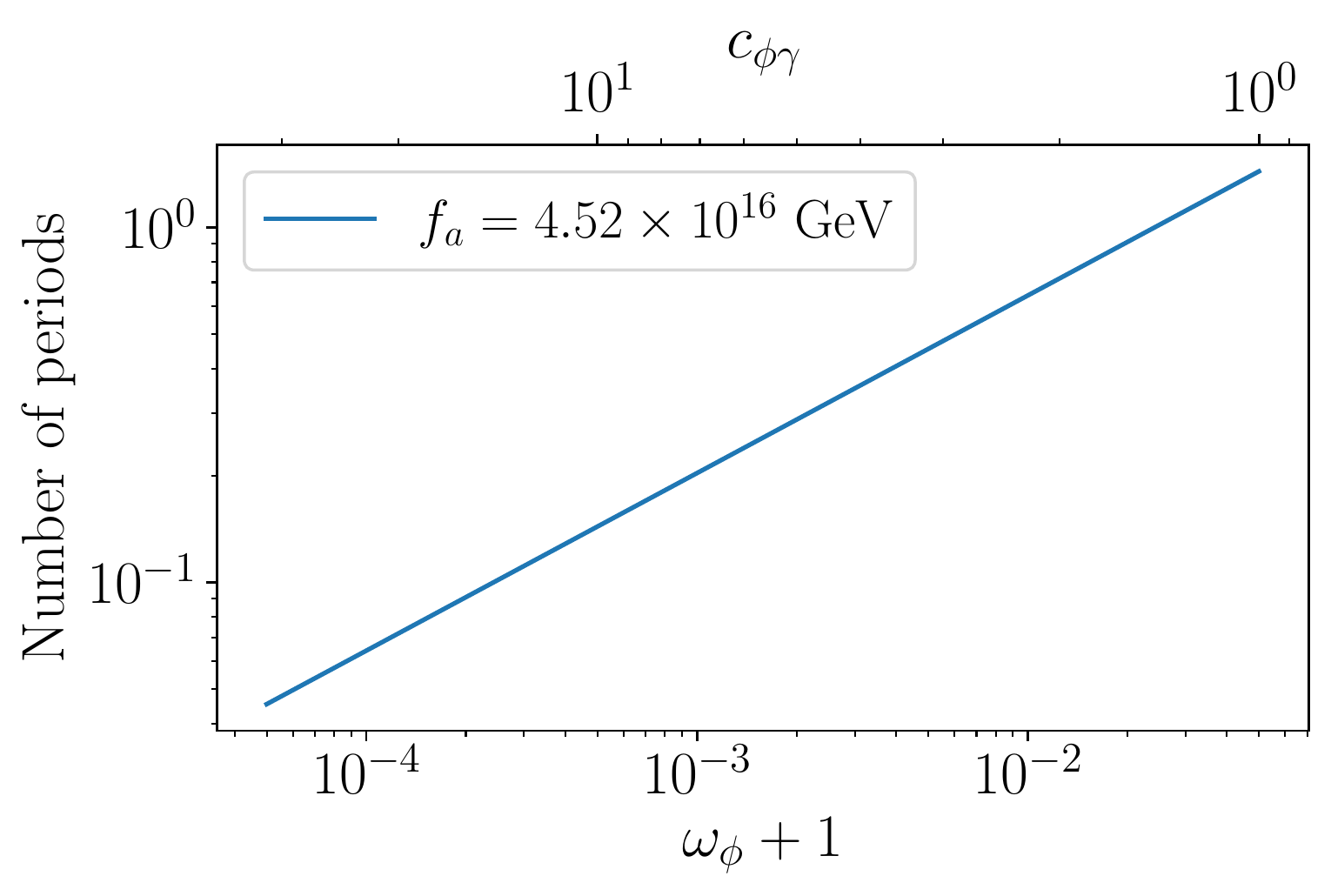}
    \end{minipage}
    \caption{On the left panel we show the potential defined in \eqref{eq:modulatedpotential} for the given decay constant and $b=0.85$. The dashed vertical lines indicate the initial and final value of the axion field that rolls down the potential from the LSS. The right panel shows the number of fundamental periods explored by the rolling axion as a function of the equation of state or the anomaly coefficient. The relation between these two is given by combining eqns \eqref{eq:Nosc1} and \eqref{eq:Nosc2}. In both panels we fix the decay constant to its largest value $f_a=4.52\times 10^{16}$.}
    \label{fig:POtwithoscillations}
\end{figure}

The number of periods explored by the axion field from the LSS is given by:
\begin{equation}\label{eq:Nosc1}
    N_{osc}=\frac{\Delta\phi}{2\pi f_a}= 
    \frac{ 2\beta}{\alpha_{em}c_{\phi\gamma}}=\frac{1.43}{c_{\phi\gamma}}\Big(\frac{\beta}{0.30\text{deg}}\Big),
\end{equation}
where eqns. \eqref{eq:betaformula} and \eqref{eq:coupling&fa} have been used. Interestingly, $N_{osc}$ only depends on $\beta$ and the axion-photon anomaly coefficient $c_{\phi\gamma}$. This is a model-dependent parameter and its natural value is of order $\mathcal{O}(1)$ \cite{Axiverse}, therefore, substituting the best fit value of $\beta$, eq. \eqref{eq:Nosc1} predicts that the field rolls over almost a period and a half. Alternatively, we can rewrite $c_{\phi\gamma}$ in terms of the decay constant and the equation of state via \eqref{eq:decayres} and we obtain \footnote{Here we fix the axion abundance to $\Omega_\phi=0.69$.}: 
\begin{equation}\label{eq:Nosc2}
    N_{osc}=1.43 \times \Big(\frac{4.52\times10^{16} \text{GeV}}{f_a}\Big)\Big(\frac{\omega_\phi+1}{0.05}\Big)^{\frac{1}{2}} \ .
\end{equation}
We learn that increasing the decay constant reduces the number of oscillations, but makes the amplitude of the oscillations larger, for given $b$, as follows from \eqref{eq:modulatedpotential}. On the other hand, a smaller equation of state gives both less oscillations and lowers their amplitudes. Therefore, the greatest contribution from the periodic potential is achieved when the decay constant and the equation of state take the largest value that are allowed from previous constraints. It follows that, for a fixed value of the decay constant, there is a one-to-one correspondence between the number of periods potentially observed in the DE and the anomaly coefficient or the equation of state as can be seen in the right of figure \ref{fig:POtwithoscillations}. We further note that, for a given $f_a$, the value of $c_{\phi\gamma}$ increases with the decrease of the equation of state that, at the same time, leads to a less evident oscillations. Remarkably, we can see from the plot in figure \ref{fig:POtwithoscillations} that $c_{\phi\gamma}$ of order unity implies $\omega_\phi+1\simeq 0.05$ when $f_a=4.52\times 10^{16}$ GeV. Finally, the possibility of observing the peculiar oscillations of monodromic DE highly depends on the value of $f_a$ and $\omega_\phi+1$ and the effect is bigger when they are close to their largest values. Therefore, looking for those oscillations with future surveys might give a chance to probe axion DE with a decay constant on the order of the GUT scale. This observation would give a strong support for the monodromy scenario. On the other hand,  no observations of such oscillations would imply that the decay constant and the equation of state are much smaller.  
       
\subsection{Evolving birefringence angle}
In this section, we discuss the late time evolution of the birefringence angle. Current analysis of {\it Planck} data only use modes $l\geq 51$ \cite{Minami:2020odp, diegopalazuelos2022cosmic,eskilt2022}, but future experiments as LiteBIRD \cite{2019JLTP..194..443H} are sensitive to $l\geq2$. In particular, for multiples $l\lesssim 20$, we can extract the polarization effect from the epoch of reionization of hydrogen atoms with a redshift $z_{rei}\sim 8$, which occurs at much later time compared to the recombination epoch $z_{rec}\sim 1090$. Namely, we could learn about the time evolution of $\beta$ by probing its $l$-dependence.
Inspired by such tomographic approach, authors of \cite{sherwin2021cosmic} have recently proposed to measure the difference of the birefringence angle between the epochs of recombination and reionization: $\Delta\beta= \beta_{rec}- \beta_{rei}$. In this regard, measuring $\Delta\beta \approx \beta_{rec}$ would point toward CB sources occurring before reionization corresponding to axions with masses $m_a\gtrsim 10^{-32}$ eV \cite{Nakatsuka:2022epj}. This includes also early DE explanations. On the other hand, measuring a much smaller $\Delta\beta \ll 1$ would support the late DE origin of the signal. In the case of late DE, we still expect a mild time-dependence of $\beta$. We can estimate this for the linear potential, since we have previously seen that, in the slow-roll regime, the field displacement is sensitive only to the local steepness of the potential. Indeed, from the linear evolution of the scalar field \eqref{eq:MDsolution} and eq. \eqref{eq:betaformula} it follows that:
\begin{equation}
    \beta\simeq \frac{g_{\phi\gamma}}{2}s\int_{\tau_{em}}^{\tau_0}\tau \mathrm{d}\tau. 
\end{equation}
The evolution of $\beta$ depends on the coupling, the slope and the time interval. Taking the ratio of $\beta_{rei}$ and $\beta_{rec}$, the dependence on the first two cancels out and we have:
\begin{equation}\label{eq:deltabeta}
    \frac{\beta_{rei}}{\beta_{rec}}\simeq \frac{\tau_{0}^2-\tau_{rei}^2}{\tau_{0}^2-\tau_{rec}^2}\simeq 1-\Big(\frac{\tau_{rei}}{\tau_0}\Big)^2,
\end{equation}
where we have neglected $\tau_{rec}\ll\tau_{rei}$. Subsequently, substituting $\tau_{rei}= \frac{2}{3\sqrt{\Omega_m}}(1+z_{rei})^{-3}$ we find that the relative difference \eqref{eq:deltabeta} is:
\begin{equation}\label{eq:deltabetabeta}
    \frac{\beta_{rec}-\beta_{rei}}{\beta_{rec}}= \frac{\Delta\beta}{\beta_{rec}}\simeq \frac{4}{9\Omega_m\tau_0^2}\frac{1}{(1+z_{rei})^3} \sim \mathcal{O}(10^{-3}).
\end{equation}
Thus, we found that the generic expectation of DE is that the relative value of the $\beta$-angles only depends on the time interval between their sources. Measuring such tiny value would be, however, challenging for future surveys.

\vskip\baselineskip
\section{Conclusion}
\label{conc}
In this work, we studied the new implications of the CB measurement for axion DE with monodromy potential. At first we analyzed the case of axion with a linear potential in terms of its slope. We found the allowed parameter space consistent with the constraints from the equation of state for DE by {\it Planck} 2018 and the axion-photon coupling constant by {\it Chandra}. Remarkably, we found that the corresponding decay constant is sub-Planckian and has an upper bound of $f_a\leq 4.52\times 10^{16} \ \text{GeV}$, which is in the favourable region predicted from string theory calculations. Subsequently, we considered a broader class of monomial potentials predicted by axion monodromies. As before, we studied the induced CB in terms of the slow-roll parameter and we restricted the corresponding parameter space. We note that the current constraints on the equation of state require a super-Planckian value of the field which can be naturally accounted by the monodromy. Using the birefringence angle, we found that the relation between the decay constant and the equation of state still holds for this broader case. Therefore, the axion monodromy can avoid the conventional issues of the standard axion DE model. Intriguingly, the birefringence measurement explained by axion DE gives the first lower bound for the equation of state, therefore to the slow-roll parameter, that has important consequences for the swampland conjectures.  
Finally, we discussed the possibility of oscillations in the DE sector from the periodic corrections by instantons to the monodromy potential. Remarkably, we found that the measurement of $\beta$ could link the number of oscillations to the value of the equation of state or the anomaly coefficient.

With future surveys we have a chance to probe further this model by looking for such oscillations and by measuring the birefringence evolution from the CMB polarization map at different multipoles. This measurement, together with a better understanding of its systematics, will clarify the cosmological significance of the signal and will have profound implications for our understanding of the underlying parity-violating physics. A further extension of this work would be to consider the effect of multiple and mixed monodromy axions as they naturally come from the ``axiverse" (see e.g. \cite{DAmico:2016jbm,Kumar:2013oda}). We also expect that possible interactions between DE and DM would have considerable consequences on the effective field displacement because of the energy transfer between those two sectors. We leave these discussions to our future works.

\begin{acknowledgments}

We would like to give spacial thanks to Eiichiro Komatsu for the fruitful discussion and helpful comments.
We are also grateful to Diego Blas and Miguel Escudero for valuable comments on the draft of the paper.
During this work, SG has been partially supported by \textit{LMU Study Completion Scholarship} and by grants
PID2020-115845GB-I00/AEI/10.13039/501100011033 and 2017-SGR-1069.
IFAE is partially funded by the CERCA program of the Generalitat de Catalunya. IO has been supported by \textit{JSPS Overseas Research Fellowship} and by \textit{JSPS KAKENHI Grant Number JP20H05859}. Some of the results of this paper were presented during the online workshop on \textit{Very Light Dark Matter 2021} organized by Kavli Institute for the Physics and Mathematics of the Universe, University of Tokyo.

\end{acknowledgments}
\bibliographystyle{apsrev4-1}
\bibliography{literatur}

\begin{thebibliography}{117}%
\makeatletter
\providecommand \@ifxundefined [1]{%
 \@ifx{#1\undefined}
}%
\providecommand \@ifnum [1]{%
 \ifnum #1\expandafter \@firstoftwo
 \else \expandafter \@secondoftwo
 \fi
}%
\providecommand \@ifx [1]{%
 \ifx #1\expandafter \@firstoftwo
 \else \expandafter \@secondoftwo
 \fi
}%
\providecommand \natexlab [1]{#1}%
\providecommand \enquote  [1]{``#1''}%
\providecommand \bibnamefont  [1]{#1}%
\providecommand \bibfnamefont [1]{#1}%
\providecommand \citenamefont [1]{#1}%
\providecommand \href@noop [0]{\@secondoftwo}%
\providecommand \href [0]{\begingroup \@sanitize@url \@href}%
\providecommand \@href[1]{\@@startlink{#1}\@@href}%
\providecommand \@@href[1]{\endgroup#1\@@endlink}%
\providecommand \@sanitize@url [0]{\catcode `\\12\catcode `\$12\catcode
  `\&12\catcode `\#12\catcode `\^12\catcode `\_12\catcode `\%12\relax}%
\providecommand \@@startlink[1]{}%
\providecommand \@@endlink[0]{}%
\providecommand \url  [0]{\begingroup\@sanitize@url \@url }%
\providecommand \@url [1]{\endgroup\@href {#1}{\urlprefix }}%
\providecommand \urlprefix  [0]{URL }%
\providecommand \Eprint [0]{\href }%
\providecommand \doibase [0]{http://dx.doi.org/}%
\providecommand \selectlanguage [0]{\@gobble}%
\providecommand \bibinfo  [0]{\@secondoftwo}%
\providecommand \bibfield  [0]{\@secondoftwo}%
\providecommand \translation [1]{[#1]}%
\providecommand \BibitemOpen [0]{}%
\providecommand \bibitemStop [0]{}%
\providecommand \bibitemNoStop [0]{.\EOS\space}%
\providecommand \EOS [0]{\spacefactor3000\relax}%
\providecommand \BibitemShut  [1]{\csname bibitem#1\endcsname}%
\let\auto@bib@innerbib\@empty
\bibitem [{\citenamefont {Weinberg}(1989)}]{CCPWeinberg}%
  \BibitemOpen
  \bibfield  {author} {\bibinfo {author} {\bibfnamefont {S.}~\bibnamefont
  {Weinberg}},\ }\href {\doibase 10.1103/RevModPhys.61.1} {\bibfield  {journal}
  {\bibinfo  {journal} {Rev. Mod. Phys.}\ }\textbf {\bibinfo {volume} {61}},\
  \bibinfo {pages} {1} (\bibinfo {year} {1989})}\BibitemShut {NoStop}%
\bibitem [{\citenamefont {Peebles}\ and\ \citenamefont
  {Ratra}(2003)}]{Peebles:2002gy}%
  \BibitemOpen
  \bibfield  {author} {\bibinfo {author} {\bibfnamefont {P.~J.~E.}\
  \bibnamefont {Peebles}}\ and\ \bibinfo {author} {\bibfnamefont
  {B.}~\bibnamefont {Ratra}},\ }\href {\doibase 10.1103/RevModPhys.75.559}
  {\bibfield  {journal} {\bibinfo  {journal} {Rev. Mod. Phys.}\ }\textbf
  {\bibinfo {volume} {75}},\ \bibinfo {pages} {559} (\bibinfo {year}
  {2003})}\BibitemShut {NoStop}%
\bibitem [{\citenamefont {Copeland}\ \emph {et~al.}(2006)\citenamefont
  {Copeland}, \citenamefont {Sami},\ and\ \citenamefont
  {Tsujikawa}}]{Copeland:2006wr}%
  \BibitemOpen
  \bibfield  {author} {\bibinfo {author} {\bibfnamefont {E.~J.}\ \bibnamefont
  {Copeland}}, \bibinfo {author} {\bibfnamefont {M.}~\bibnamefont {Sami}}, \
  and\ \bibinfo {author} {\bibfnamefont {S.}~\bibnamefont {Tsujikawa}},\ }\href
  {\doibase 10.1142/S021827180600942X} {\bibfield  {journal} {\bibinfo
  {journal} {Int. J. Mod. Phys. D}\ }\textbf {\bibinfo {volume} {15}},\
  \bibinfo {pages} {1753} (\bibinfo {year} {2006})},\ \Eprint
  {http://arxiv.org/abs/hep-th/0603057} {arXiv:hep-th/0603057} \BibitemShut
  {NoStop}%
\bibitem [{\citenamefont {Tsujikawa}(2011)}]{Tsujikawa:2010sc}%
  \BibitemOpen
  \bibfield  {author} {\bibinfo {author} {\bibfnamefont {S.}~\bibnamefont
  {Tsujikawa}},\ }\href {\doibase 10.1007/978-90-481-8685-3_8} {\bibfield
  {journal} {\bibinfo  {journal} {Astrophysics and Space Science Library}\ ,\
  \bibinfo {pages} {331–402}} (\bibinfo {year} {2011})}\BibitemShut {NoStop}%
\bibitem [{\citenamefont {Peccei}\ and\ \citenamefont
  {Quinn}(1977)}]{Peccei:1977hh}%
  \BibitemOpen
  \bibfield  {author} {\bibinfo {author} {\bibfnamefont {R.~D.}\ \bibnamefont
  {Peccei}}\ and\ \bibinfo {author} {\bibfnamefont {H.~R.}\ \bibnamefont
  {Quinn}},\ }\href {\doibase 10.1103/PhysRevLett.38.1440} {\bibfield
  {journal} {\bibinfo  {journal} {Phys. Rev. Lett.}\ }\textbf {\bibinfo
  {volume} {38}},\ \bibinfo {pages} {1440} (\bibinfo {year}
  {1977})}\BibitemShut {NoStop}%
\bibitem [{\citenamefont {Weinberg}(1978)}]{Weinberg:1977ma}%
  \BibitemOpen
  \bibfield  {author} {\bibinfo {author} {\bibfnamefont {S.}~\bibnamefont
  {Weinberg}},\ }\href {\doibase 10.1103/PhysRevLett.40.223} {\bibfield
  {journal} {\bibinfo  {journal} {Phys. Rev. Lett.}\ }\textbf {\bibinfo
  {volume} {40}},\ \bibinfo {pages} {223} (\bibinfo {year} {1978})}\BibitemShut
  {NoStop}%
\bibitem [{\citenamefont {Wilczek}(1978)}]{Wilczek:1977pj}%
  \BibitemOpen
  \bibfield  {author} {\bibinfo {author} {\bibfnamefont {F.}~\bibnamefont
  {Wilczek}},\ }\href {\doibase 10.1103/PhysRevLett.40.279} {\bibfield
  {journal} {\bibinfo  {journal} {Phys. Rev. Lett.}\ }\textbf {\bibinfo
  {volume} {40}},\ \bibinfo {pages} {279} (\bibinfo {year} {1978})}\BibitemShut
  {NoStop}%
\bibitem [{\citenamefont {Frieman}\ \emph {et~al.}(1995)\citenamefont
  {Frieman}, \citenamefont {Hill}, \citenamefont {Stebbins},\ and\
  \citenamefont {Waga}}]{Frieman_1995}%
  \BibitemOpen
  \bibfield  {author} {\bibinfo {author} {\bibfnamefont {J.~A.}\ \bibnamefont
  {Frieman}}, \bibinfo {author} {\bibfnamefont {C.~T.}\ \bibnamefont {Hill}},
  \bibinfo {author} {\bibfnamefont {A.}~\bibnamefont {Stebbins}}, \ and\
  \bibinfo {author} {\bibfnamefont {I.}~\bibnamefont {Waga}},\ }\href {\doibase
  10.1103/PhysRevLett.75.2077} {\bibfield  {journal} {\bibinfo  {journal}
  {Phys. Rev. Lett.}\ }\textbf {\bibinfo {volume} {75}},\ \bibinfo {pages}
  {2077} (\bibinfo {year} {1995})},\ \Eprint
  {http://arxiv.org/abs/astro-ph/9505060} {arXiv:astro-ph/9505060} \BibitemShut
  {NoStop}%
\bibitem [{\citenamefont {Kolda}\ and\ \citenamefont
  {Lyth}(1999)}]{Kolda_1999}%
  \BibitemOpen
  \bibfield  {author} {\bibinfo {author} {\bibfnamefont {C.~F.}\ \bibnamefont
  {Kolda}}\ and\ \bibinfo {author} {\bibfnamefont {D.~H.}\ \bibnamefont
  {Lyth}},\ }\href {\doibase 10.1016/S0370-2693(99)00657-7} {\bibfield
  {journal} {\bibinfo  {journal} {Phys. Lett. B}\ }\textbf {\bibinfo {volume}
  {458}},\ \bibinfo {pages} {197} (\bibinfo {year} {1999})},\ \Eprint
  {http://arxiv.org/abs/hep-ph/9811375} {arXiv:hep-ph/9811375} \BibitemShut
  {NoStop}%
\bibitem [{\citenamefont {Svrcek}(2006)}]{Svreck2006}%
  \BibitemOpen
  \bibfield  {author} {\bibinfo {author} {\bibfnamefont {P.}~\bibnamefont
  {Svrcek}},\ }\href@noop {} {\  (\bibinfo {year} {2006})},\ \Eprint
  {http://arxiv.org/abs/hep-th/0607086} {arXiv:hep-th/0607086} \BibitemShut
  {NoStop}%
\bibitem [{\citenamefont {Svrcek}\ and\ \citenamefont
  {Witten}(2006)}]{2006JHEP...06..051S}%
  \BibitemOpen
  \bibfield  {author} {\bibinfo {author} {\bibfnamefont {P.}~\bibnamefont
  {Svrcek}}\ and\ \bibinfo {author} {\bibfnamefont {E.}~\bibnamefont
  {Witten}},\ }\href {\doibase 10.1088/1126-6708/2006/06/051} {\bibfield
  {journal} {\bibinfo  {journal} {Journal of High Energy Physics}\ }\textbf
  {\bibinfo {volume} {2006}},\ \bibinfo {pages} {051–051} (\bibinfo {year}
  {2006})}\BibitemShut {NoStop}%
\bibitem [{\citenamefont {Arvanitaki}\ \emph {et~al.}(2010)\citenamefont
  {Arvanitaki}, \citenamefont {Dimopoulos}, \citenamefont {Dubovsky},
  \citenamefont {Kaloper},\ and\ \citenamefont {March-Russell}}]{Axiverse}%
  \BibitemOpen
  \bibfield  {author} {\bibinfo {author} {\bibfnamefont {A.}~\bibnamefont
  {Arvanitaki}}, \bibinfo {author} {\bibfnamefont {S.}~\bibnamefont
  {Dimopoulos}}, \bibinfo {author} {\bibfnamefont {S.}~\bibnamefont
  {Dubovsky}}, \bibinfo {author} {\bibfnamefont {N.}~\bibnamefont {Kaloper}}, \
  and\ \bibinfo {author} {\bibfnamefont {J.}~\bibnamefont {March-Russell}},\
  }\href {\doibase 10.1103/PhysRevD.81.123530} {\bibfield  {journal} {\bibinfo
  {journal} {Phys. Rev. D}\ }\textbf {\bibinfo {volume} {81}},\ \bibinfo
  {pages} {123530} (\bibinfo {year} {2010})},\ \Eprint
  {http://arxiv.org/abs/0905.4720} {arXiv:0905.4720 [hep-th]} \BibitemShut
  {NoStop}%
\bibitem [{\citenamefont {Kim}(1999)}]{Kim1999}%
  \BibitemOpen
  \bibfield  {author} {\bibinfo {author} {\bibfnamefont {J.~E.}\ \bibnamefont
  {Kim}},\ }\href {\doibase 10.1088/1126-6708/1999/05/022} {\bibfield
  {journal} {\bibinfo  {journal} {JHEP}\ }\textbf {\bibinfo {volume} {05}},\
  \bibinfo {pages} {022} (\bibinfo {year} {1999})},\ \Eprint
  {http://arxiv.org/abs/hep-ph/9811509} {arXiv:hep-ph/9811509} \BibitemShut
  {NoStop}%
\bibitem [{\citenamefont {Kim}\ and\ \citenamefont {Nilles}(2003)}]{Kim_2003}%
  \BibitemOpen
  \bibfield  {author} {\bibinfo {author} {\bibfnamefont {J.~E.}\ \bibnamefont
  {Kim}}\ and\ \bibinfo {author} {\bibfnamefont {H.~P.}\ \bibnamefont
  {Nilles}},\ }\href {\doibase 10.1016/S0370-2693(02)03148-9} {\bibfield
  {journal} {\bibinfo  {journal} {Phys. Lett. B}\ }\textbf {\bibinfo {volume}
  {553}},\ \bibinfo {pages} {1} (\bibinfo {year} {2003})},\ \Eprint
  {http://arxiv.org/abs/hep-ph/0210402} {arXiv:hep-ph/0210402} \BibitemShut
  {NoStop}%
\bibitem [{\citenamefont {Choi}(2000)}]{Choi2000}%
  \BibitemOpen
  \bibfield  {author} {\bibinfo {author} {\bibfnamefont {K.}~\bibnamefont
  {Choi}},\ }\href {\doibase 10.1103/PhysRevD.62.043509} {\bibfield  {journal}
  {\bibinfo  {journal} {Phys. Rev. D}\ }\textbf {\bibinfo {volume} {62}},\
  \bibinfo {pages} {043509} (\bibinfo {year} {2000})},\ \Eprint
  {http://arxiv.org/abs/hep-ph/9902292} {arXiv:hep-ph/9902292} \BibitemShut
  {NoStop}%
\bibitem [{\citenamefont {Kim}\ and\ \citenamefont {Nilles}(2009)}]{Kim2009}%
  \BibitemOpen
  \bibfield  {author} {\bibinfo {author} {\bibfnamefont {J.~E.}\ \bibnamefont
  {Kim}}\ and\ \bibinfo {author} {\bibfnamefont {H.~P.}\ \bibnamefont
  {Nilles}},\ }\href {\doibase 10.1088/1475-7516/2009/05/010} {\bibfield
  {journal} {\bibinfo  {journal} {Journal of Cosmology and Astroparticle
  Physics}\ }\textbf {\bibinfo {volume} {2009}},\ \bibinfo {pages} {010–010}
  (\bibinfo {year} {2009})}\BibitemShut {NoStop}%
\bibitem [{\citenamefont {Marsh}(2016)}]{Marsh:2015xka}%
  \BibitemOpen
  \bibfield  {author} {\bibinfo {author} {\bibfnamefont {D.~J.~E.}\
  \bibnamefont {Marsh}},\ }\href {\doibase 10.1016/j.physrep.2016.06.005}
  {\bibfield  {journal} {\bibinfo  {journal} {Phys. Rept.}\ }\textbf {\bibinfo
  {volume} {643}},\ \bibinfo {pages} {1} (\bibinfo {year} {2016})},\ \Eprint
  {http://arxiv.org/abs/1510.07633} {arXiv:1510.07633 [astro-ph.CO]}
  \BibitemShut {NoStop}%
\bibitem [{\citenamefont {Sikivie}(1983)}]{Sikivie:1983ip}%
  \BibitemOpen
  \bibfield  {author} {\bibinfo {author} {\bibfnamefont {P.}~\bibnamefont
  {Sikivie}},\ }\href {\doibase 10.1103/PhysRevLett.51.1415} {\bibfield
  {journal} {\bibinfo  {journal} {Phys. Rev. Lett.}\ }\textbf {\bibinfo
  {volume} {51}},\ \bibinfo {pages} {1415} (\bibinfo {year} {1983})},\ \bibinfo
  {note} {[Erratum: Phys.Rev.Lett. 52, 695 (1984)]}\BibitemShut {NoStop}%
\bibitem [{\citenamefont {Raffelt}\ and\ \citenamefont
  {Stodolsky}(1988)}]{Raffelt:1987im}%
  \BibitemOpen
  \bibfield  {author} {\bibinfo {author} {\bibfnamefont {G.}~\bibnamefont
  {Raffelt}}\ and\ \bibinfo {author} {\bibfnamefont {L.}~\bibnamefont
  {Stodolsky}},\ }\href {\doibase 10.1103/PhysRevD.37.1237} {\bibfield
  {journal} {\bibinfo  {journal} {Phys. Rev. D}\ }\textbf {\bibinfo {volume}
  {37}},\ \bibinfo {pages} {1237} (\bibinfo {year} {1988})}\BibitemShut
  {NoStop}%
\bibitem [{\citenamefont {Carroll}\ \emph {et~al.}(1990)\citenamefont
  {Carroll}, \citenamefont {Field},\ and\ \citenamefont
  {Jackiw}}]{Carroll:1989vb}%
  \BibitemOpen
  \bibfield  {author} {\bibinfo {author} {\bibfnamefont {S.~M.}\ \bibnamefont
  {Carroll}}, \bibinfo {author} {\bibfnamefont {G.~B.}\ \bibnamefont {Field}},
  \ and\ \bibinfo {author} {\bibfnamefont {R.}~\bibnamefont {Jackiw}},\ }\href
  {\doibase 10.1103/PhysRevD.41.1231} {\bibfield  {journal} {\bibinfo
  {journal} {Phys. Rev. D}\ }\textbf {\bibinfo {volume} {41}},\ \bibinfo
  {pages} {1231} (\bibinfo {year} {1990})}\BibitemShut {NoStop}%
\bibitem [{\citenamefont {Carroll}\ and\ \citenamefont
  {Field}(1991)}]{Carroll:1991zs}%
  \BibitemOpen
  \bibfield  {author} {\bibinfo {author} {\bibfnamefont {S.~M.}\ \bibnamefont
  {Carroll}}\ and\ \bibinfo {author} {\bibfnamefont {G.~B.}\ \bibnamefont
  {Field}},\ }\href {\doibase 10.1103/PhysRevD.43.3789} {\bibfield  {journal}
  {\bibinfo  {journal} {Phys. Rev. D}\ }\textbf {\bibinfo {volume} {43}},\
  \bibinfo {pages} {3789} (\bibinfo {year} {1991})}\BibitemShut {NoStop}%
\bibitem [{\citenamefont {Harari}\ and\ \citenamefont
  {Sikivie}(1992)}]{Harari:1992ea}%
  \BibitemOpen
  \bibfield  {author} {\bibinfo {author} {\bibfnamefont {D.}~\bibnamefont
  {Harari}}\ and\ \bibinfo {author} {\bibfnamefont {P.}~\bibnamefont
  {Sikivie}},\ }\href {\doibase 10.1016/0370-2693(92)91363-E} {\bibfield
  {journal} {\bibinfo  {journal} {Phys. Lett. B}\ }\textbf {\bibinfo {volume}
  {289}},\ \bibinfo {pages} {67} (\bibinfo {year} {1992})}\BibitemShut
  {NoStop}%
\bibitem [{\citenamefont {Carroll}(1998)}]{Carroll:1998zi}%
  \BibitemOpen
  \bibfield  {author} {\bibinfo {author} {\bibfnamefont {S.~M.}\ \bibnamefont
  {Carroll}},\ }\href {\doibase 10.1103/PhysRevLett.81.3067} {\bibfield
  {journal} {\bibinfo  {journal} {Phys. Rev. Lett.}\ }\textbf {\bibinfo
  {volume} {81}},\ \bibinfo {pages} {3067} (\bibinfo {year} {1998})},\ \Eprint
  {http://arxiv.org/abs/astro-ph/9806099} {arXiv:astro-ph/9806099} \BibitemShut
  {NoStop}%
\bibitem [{\citenamefont {Komatsu}(2022)}]{komatsu2022new}%
  \BibitemOpen
  \bibfield  {author} {\bibinfo {author} {\bibfnamefont {E.}~\bibnamefont
  {Komatsu}},\ }\href@noop {} {\  (\bibinfo {year} {2022})},\ \Eprint
  {http://arxiv.org/abs/2202.13919} {arXiv:2202.13919 [astro-ph.CO]}
  \BibitemShut {NoStop}%
\bibitem [{\citenamefont {Lue}\ \emph {et~al.}(1999)\citenamefont {Lue},
  \citenamefont {Wang},\ and\ \citenamefont {Kamionkowski}}]{Lue:1998mq}%
  \BibitemOpen
  \bibfield  {author} {\bibinfo {author} {\bibfnamefont {A.}~\bibnamefont
  {Lue}}, \bibinfo {author} {\bibfnamefont {L.-M.}\ \bibnamefont {Wang}}, \
  and\ \bibinfo {author} {\bibfnamefont {M.}~\bibnamefont {Kamionkowski}},\
  }\href {\doibase 10.1103/PhysRevLett.83.1506} {\bibfield  {journal} {\bibinfo
   {journal} {Phys. Rev. Lett.}\ }\textbf {\bibinfo {volume} {83}},\ \bibinfo
  {pages} {1506} (\bibinfo {year} {1999})},\ \Eprint
  {http://arxiv.org/abs/astro-ph/9812088} {arXiv:astro-ph/9812088} \BibitemShut
  {NoStop}%
\bibitem [{\citenamefont {Wu}\ \emph {et~al.}(2009)\citenamefont {Wu} \emph
  {et~al.}}]{QUaD:2008ado}%
  \BibitemOpen
  \bibfield  {author} {\bibinfo {author} {\bibfnamefont {E.~Y.~S.}\
  \bibnamefont {Wu}} \emph {et~al.} (\bibinfo {collaboration} {QUaD}),\ }\href
  {\doibase 10.1103/PhysRevLett.102.161302} {\bibfield  {journal} {\bibinfo
  {journal} {Phys. Rev. Lett.}\ }\textbf {\bibinfo {volume} {102}},\ \bibinfo
  {pages} {161302} (\bibinfo {year} {2009})},\ \Eprint
  {http://arxiv.org/abs/0811.0618} {arXiv:0811.0618 [astro-ph]} \BibitemShut
  {NoStop}%
\bibitem [{\citenamefont {Komatsu}\ \emph {et~al.}(2011)\citenamefont
  {Komatsu}, \citenamefont {Smith}, \citenamefont {Dunkley}, \citenamefont
  {Bennett}, \citenamefont {Gold}, \citenamefont {Hinshaw}, \citenamefont
  {Jarosik}, \citenamefont {Larson}, \citenamefont {Nolta}, \citenamefont
  {Page}, \citenamefont {Spergel}, \citenamefont {Halpern}, \citenamefont
  {Hill}, \citenamefont {Kogut}, \citenamefont {Limon}, \citenamefont {Meyer},
  \citenamefont {Odegard}, \citenamefont {Tucker}, \citenamefont {Weiland},
  \citenamefont {Wollack},\ and\ \citenamefont {Wright}}]{Komatsu_2011}%
  \BibitemOpen
  \bibfield  {author} {\bibinfo {author} {\bibfnamefont {E.}~\bibnamefont
  {Komatsu}}, \bibinfo {author} {\bibfnamefont {K.~M.}\ \bibnamefont {Smith}},
  \bibinfo {author} {\bibfnamefont {J.}~\bibnamefont {Dunkley}}, \bibinfo
  {author} {\bibfnamefont {C.~L.}\ \bibnamefont {Bennett}}, \bibinfo {author}
  {\bibfnamefont {B.}~\bibnamefont {Gold}}, \bibinfo {author} {\bibfnamefont
  {G.}~\bibnamefont {Hinshaw}}, \bibinfo {author} {\bibfnamefont
  {N.}~\bibnamefont {Jarosik}}, \bibinfo {author} {\bibfnamefont
  {D.}~\bibnamefont {Larson}}, \bibinfo {author} {\bibfnamefont {M.~R.}\
  \bibnamefont {Nolta}}, \bibinfo {author} {\bibfnamefont {L.}~\bibnamefont
  {Page}}, \bibinfo {author} {\bibfnamefont {D.~N.}\ \bibnamefont {Spergel}},
  \bibinfo {author} {\bibfnamefont {M.}~\bibnamefont {Halpern}}, \bibinfo
  {author} {\bibfnamefont {R.~S.}\ \bibnamefont {Hill}}, \bibinfo {author}
  {\bibfnamefont {A.}~\bibnamefont {Kogut}}, \bibinfo {author} {\bibfnamefont
  {M.}~\bibnamefont {Limon}}, \bibinfo {author} {\bibfnamefont {S.~S.}\
  \bibnamefont {Meyer}}, \bibinfo {author} {\bibfnamefont {N.}~\bibnamefont
  {Odegard}}, \bibinfo {author} {\bibfnamefont {G.~S.}\ \bibnamefont {Tucker}},
  \bibinfo {author} {\bibfnamefont {J.~L.}\ \bibnamefont {Weiland}}, \bibinfo
  {author} {\bibfnamefont {E.}~\bibnamefont {Wollack}}, \ and\ \bibinfo
  {author} {\bibfnamefont {E.~L.}\ \bibnamefont {Wright}},\ }\href {\doibase
  10.1088/0067-0049/192/2/18} {\bibfield  {journal} {\bibinfo  {journal} {The
  Astrophysical Journal Supplement Series}\ }\textbf {\bibinfo {volume}
  {192}},\ \bibinfo {pages} {18} (\bibinfo {year} {2011})}\BibitemShut
  {NoStop}%
\bibitem [{\citenamefont {Minami}\ \emph {et~al.}(2019)\citenamefont {Minami},
  \citenamefont {Ochi}, \citenamefont {Ichiki}, \citenamefont {Katayama},
  \citenamefont {Komatsu},\ and\ \citenamefont
  {Matsumura}}]{minami2019simultaneous}%
  \BibitemOpen
  \bibfield  {author} {\bibinfo {author} {\bibfnamefont {Y.}~\bibnamefont
  {Minami}}, \bibinfo {author} {\bibfnamefont {H.}~\bibnamefont {Ochi}},
  \bibinfo {author} {\bibfnamefont {K.}~\bibnamefont {Ichiki}}, \bibinfo
  {author} {\bibfnamefont {N.}~\bibnamefont {Katayama}}, \bibinfo {author}
  {\bibfnamefont {E.}~\bibnamefont {Komatsu}}, \ and\ \bibinfo {author}
  {\bibfnamefont {T.}~\bibnamefont {Matsumura}},\ }\href {\doibase
  10.1093/ptep/ptz079} {\bibfield  {journal} {\bibinfo  {journal} {PTEP}\
  }\textbf {\bibinfo {volume} {2019}},\ \bibinfo {pages} {083E02} (\bibinfo
  {year} {2019})},\ \Eprint {http://arxiv.org/abs/1904.12440} {arXiv:1904.12440
  [astro-ph.CO]} \BibitemShut {NoStop}%
\bibitem [{\citenamefont {Minami}\ and\ \citenamefont
  {Komatsu}(2020)}]{Minami:2020odp}%
  \BibitemOpen
  \bibfield  {author} {\bibinfo {author} {\bibfnamefont {Y.}~\bibnamefont
  {Minami}}\ and\ \bibinfo {author} {\bibfnamefont {E.}~\bibnamefont
  {Komatsu}},\ }\href {\doibase 10.1103/PhysRevLett.125.221301} {\bibfield
  {journal} {\bibinfo  {journal} {Phys. Rev. Lett.}\ }\textbf {\bibinfo
  {volume} {125}},\ \bibinfo {pages} {221301} (\bibinfo {year} {2020})},\
  \Eprint {http://arxiv.org/abs/2011.11254} {arXiv:2011.11254 [astro-ph.CO]}
  \BibitemShut {NoStop}%
\bibitem [{\citenamefont {Fujita}\ \emph
  {et~al.}(2021{\natexlab{a}})\citenamefont {Fujita}, \citenamefont {Minami},
  \citenamefont {Murai},\ and\ \citenamefont {Nakatsuka}}]{Fujita_2021}%
  \BibitemOpen
  \bibfield  {author} {\bibinfo {author} {\bibfnamefont {T.}~\bibnamefont
  {Fujita}}, \bibinfo {author} {\bibfnamefont {Y.}~\bibnamefont {Minami}},
  \bibinfo {author} {\bibfnamefont {K.}~\bibnamefont {Murai}}, \ and\ \bibinfo
  {author} {\bibfnamefont {H.}~\bibnamefont {Nakatsuka}},\ }\href {\doibase
  10.1103/PhysRevD.103.063508} {\bibfield  {journal} {\bibinfo  {journal}
  {Phys. Rev. D}\ }\textbf {\bibinfo {volume} {103}},\ \bibinfo {pages}
  {063508} (\bibinfo {year} {2021}{\natexlab{a}})},\ \Eprint
  {http://arxiv.org/abs/2008.02473} {arXiv:2008.02473 [astro-ph.CO]}
  \BibitemShut {NoStop}%
\bibitem [{\citenamefont {Fujita}\ \emph
  {et~al.}(2021{\natexlab{b}})\citenamefont {Fujita}, \citenamefont {Murai},
  \citenamefont {Nakatsuka},\ and\ \citenamefont {Tsujikawa}}]{FujitaTOP}%
  \BibitemOpen
  \bibfield  {author} {\bibinfo {author} {\bibfnamefont {T.}~\bibnamefont
  {Fujita}}, \bibinfo {author} {\bibfnamefont {K.}~\bibnamefont {Murai}},
  \bibinfo {author} {\bibfnamefont {H.}~\bibnamefont {Nakatsuka}}, \ and\
  \bibinfo {author} {\bibfnamefont {S.}~\bibnamefont {Tsujikawa}},\ }\href
  {\doibase 10.1103/PhysRevD.103.043509} {\bibfield  {journal} {\bibinfo
  {journal} {Phys. Rev. D}\ }\textbf {\bibinfo {volume} {103}},\ \bibinfo
  {pages} {043509} (\bibinfo {year} {2021}{\natexlab{b}})},\ \Eprint
  {http://arxiv.org/abs/2011.11894} {arXiv:2011.11894 [astro-ph.CO]}
  \BibitemShut {NoStop}%
\bibitem [{\citenamefont {Takahashi}\ and\ \citenamefont
  {Yin}(2021)}]{Takahashi_2021}%
  \BibitemOpen
  \bibfield  {author} {\bibinfo {author} {\bibfnamefont {F.}~\bibnamefont
  {Takahashi}}\ and\ \bibinfo {author} {\bibfnamefont {W.}~\bibnamefont
  {Yin}},\ }\href {\doibase 10.1088/1475-7516/2021/04/007} {\bibfield
  {journal} {\bibinfo  {journal} {JCAP}\ }\textbf {\bibinfo {volume} {04}},\
  \bibinfo {pages} {007} (\bibinfo {year} {2021})},\ \Eprint
  {http://arxiv.org/abs/2012.11576} {arXiv:2012.11576 [hep-ph]} \BibitemShut
  {NoStop}%
\bibitem [{\citenamefont {Nagata}\ and\ \citenamefont
  {Namikawa}(2021)}]{Nagata:2021pvc}%
  \BibitemOpen
  \bibfield  {author} {\bibinfo {author} {\bibfnamefont {R.}~\bibnamefont
  {Nagata}}\ and\ \bibinfo {author} {\bibfnamefont {T.}~\bibnamefont
  {Namikawa}},\ }\href {\doibase 10.1093/ptep/ptab040} {\bibfield  {journal}
  {\bibinfo  {journal} {PTEP}\ }\textbf {\bibinfo {volume} {2021}},\ \bibinfo
  {pages} {053} (\bibinfo {year} {2021})},\ \Eprint
  {http://arxiv.org/abs/2102.00133} {arXiv:2102.00133 [astro-ph.CO]}
  \BibitemShut {NoStop}%
\bibitem [{\citenamefont {Fung}\ \emph
  {et~al.}(2021{\natexlab{a}})\citenamefont {Fung}, \citenamefont {Li},
  \citenamefont {Liu}, \citenamefont {Luu}, \citenamefont {Qiu},\ and\
  \citenamefont {Tye}}]{Fung:2021wbz}%
  \BibitemOpen
  \bibfield  {author} {\bibinfo {author} {\bibfnamefont {L.~W.~H.}\
  \bibnamefont {Fung}}, \bibinfo {author} {\bibfnamefont {L.}~\bibnamefont
  {Li}}, \bibinfo {author} {\bibfnamefont {T.}~\bibnamefont {Liu}}, \bibinfo
  {author} {\bibfnamefont {H.~N.}\ \bibnamefont {Luu}}, \bibinfo {author}
  {\bibfnamefont {Y.-C.}\ \bibnamefont {Qiu}}, \ and\ \bibinfo {author}
  {\bibfnamefont {S.~H.~H.}\ \bibnamefont {Tye}},\ }\href {\doibase
  10.1088/1475-7516/2021/08/057} {\bibfield  {journal} {\bibinfo  {journal}
  {JCAP}\ }\textbf {\bibinfo {volume} {08}},\ \bibinfo {pages} {057} (\bibinfo
  {year} {2021}{\natexlab{a}})},\ \Eprint {http://arxiv.org/abs/2102.11257}
  {arXiv:2102.11257 [hep-ph]} \BibitemShut {NoStop}%
\bibitem [{\citenamefont {Mehta}\ \emph {et~al.}(2021)\citenamefont {Mehta},
  \citenamefont {Demirtas}, \citenamefont {Long}, \citenamefont {Marsh},
  \citenamefont {McAllister},\ and\ \citenamefont {Stott}}]{Mehta:2021pwf}%
  \BibitemOpen
  \bibfield  {author} {\bibinfo {author} {\bibfnamefont {V.~M.}\ \bibnamefont
  {Mehta}}, \bibinfo {author} {\bibfnamefont {M.}~\bibnamefont {Demirtas}},
  \bibinfo {author} {\bibfnamefont {C.}~\bibnamefont {Long}}, \bibinfo {author}
  {\bibfnamefont {D.~J.~E.}\ \bibnamefont {Marsh}}, \bibinfo {author}
  {\bibfnamefont {L.}~\bibnamefont {McAllister}}, \ and\ \bibinfo {author}
  {\bibfnamefont {M.~J.}\ \bibnamefont {Stott}},\ }\href {\doibase
  10.1088/1475-7516/2021/07/033} {\bibfield  {journal} {\bibinfo  {journal}
  {JCAP}\ }\textbf {\bibinfo {volume} {07}},\ \bibinfo {pages} {033} (\bibinfo
  {year} {2021})},\ \Eprint {http://arxiv.org/abs/2103.06812} {arXiv:2103.06812
  [hep-th]} \BibitemShut {NoStop}%
\bibitem [{\citenamefont {Nakagawa}\ \emph {et~al.}(2021)\citenamefont
  {Nakagawa}, \citenamefont {Takahashi},\ and\ \citenamefont
  {Yamada}}]{CBDM2021}%
  \BibitemOpen
  \bibfield  {author} {\bibinfo {author} {\bibfnamefont {S.}~\bibnamefont
  {Nakagawa}}, \bibinfo {author} {\bibfnamefont {F.}~\bibnamefont {Takahashi}},
  \ and\ \bibinfo {author} {\bibfnamefont {M.}~\bibnamefont {Yamada}},\ }\href
  {\doibase 10.1103/PhysRevLett.127.181103} {\bibfield  {journal} {\bibinfo
  {journal} {Phys. Rev. Lett.}\ }\textbf {\bibinfo {volume} {127}},\ \bibinfo
  {pages} {181103} (\bibinfo {year} {2021})},\ \Eprint
  {http://arxiv.org/abs/2103.08153} {arXiv:2103.08153 [hep-ph]} \BibitemShut
  {NoStop}%
\bibitem [{\citenamefont {Jain}\ \emph {et~al.}(2021)\citenamefont {Jain},
  \citenamefont {Long},\ and\ \citenamefont {Amin}}]{Jain_2021}%
  \BibitemOpen
  \bibfield  {author} {\bibinfo {author} {\bibfnamefont {M.}~\bibnamefont
  {Jain}}, \bibinfo {author} {\bibfnamefont {A.~J.}\ \bibnamefont {Long}}, \
  and\ \bibinfo {author} {\bibfnamefont {M.~A.}\ \bibnamefont {Amin}},\ }\href
  {\doibase 10.1088/1475-7516/2021/05/055} {\bibfield  {journal} {\bibinfo
  {journal} {JCAP}\ }\textbf {\bibinfo {volume} {05}},\ \bibinfo {pages} {055}
  (\bibinfo {year} {2021})},\ \Eprint {http://arxiv.org/abs/2103.10962}
  {arXiv:2103.10962 [astro-ph.CO]} \BibitemShut {NoStop}%
\bibitem [{\citenamefont {Clark}\ \emph {et~al.}(2021)\citenamefont {Clark},
  \citenamefont {Kim}, \citenamefont {Hill},\ and\ \citenamefont
  {Hensley}}]{Clark:2021kze}%
  \BibitemOpen
  \bibfield  {author} {\bibinfo {author} {\bibfnamefont {S.~E.}\ \bibnamefont
  {Clark}}, \bibinfo {author} {\bibfnamefont {C.-G.}\ \bibnamefont {Kim}},
  \bibinfo {author} {\bibfnamefont {J.~C.}\ \bibnamefont {Hill}}, \ and\
  \bibinfo {author} {\bibfnamefont {B.~S.}\ \bibnamefont {Hensley}},\ }\href
  {\doibase 10.3847/1538-4357/ac0e35} {\bibfield  {journal} {\bibinfo
  {journal} {Astrophys. J.}\ }\textbf {\bibinfo {volume} {919}},\ \bibinfo
  {pages} {53} (\bibinfo {year} {2021})},\ \Eprint
  {http://arxiv.org/abs/2105.00120} {arXiv:2105.00120 [astro-ph.GA]}
  \BibitemShut {NoStop}%
\bibitem [{\citenamefont {Fung}\ \emph
  {et~al.}(2021{\natexlab{b}})\citenamefont {Fung}, \citenamefont {Li},
  \citenamefont {Liu}, \citenamefont {Luu}, \citenamefont {Qiu},\ and\
  \citenamefont {Tye}}]{Fung:2021fcj}%
  \BibitemOpen
  \bibfield  {author} {\bibinfo {author} {\bibfnamefont {L.~W.}\ \bibnamefont
  {Fung}}, \bibinfo {author} {\bibfnamefont {L.}~\bibnamefont {Li}}, \bibinfo
  {author} {\bibfnamefont {T.}~\bibnamefont {Liu}}, \bibinfo {author}
  {\bibfnamefont {H.~N.}\ \bibnamefont {Luu}}, \bibinfo {author} {\bibfnamefont
  {Y.-C.}\ \bibnamefont {Qiu}}, \ and\ \bibinfo {author} {\bibfnamefont
  {S.~H.~H.}\ \bibnamefont {Tye}},\ }\href@noop {} {\  (\bibinfo {year}
  {2021}{\natexlab{b}})},\ \Eprint {http://arxiv.org/abs/2105.01631}
  {arXiv:2105.01631 [astro-ph.CO]} \BibitemShut {NoStop}%
\bibitem [{\citenamefont {Namikawa}(2021)}]{Namikawa:2021gbr}%
  \BibitemOpen
  \bibfield  {author} {\bibinfo {author} {\bibfnamefont {T.}~\bibnamefont
  {Namikawa}},\ }\href {\doibase 10.1093/mnras/stab1796} {\bibfield  {journal}
  {\bibinfo  {journal} {Mon. Not. Roy. Astron. Soc.}\ }\textbf {\bibinfo
  {volume} {506}},\ \bibinfo {pages} {1250} (\bibinfo {year} {2021})},\ \Eprint
  {http://arxiv.org/abs/2105.03367} {arXiv:2105.03367 [astro-ph.CO]}
  \BibitemShut {NoStop}%
\bibitem [{\citenamefont {Alvey}\ and\ \citenamefont
  {Escudero~Abenza}(2021)}]{Alvey2021}%
  \BibitemOpen
  \bibfield  {author} {\bibinfo {author} {\bibfnamefont {J.}~\bibnamefont
  {Alvey}}\ and\ \bibinfo {author} {\bibfnamefont {M.}~\bibnamefont
  {Escudero~Abenza}},\ }\href {\doibase 10.1016/j.physletb.2021.136752}
  {\bibfield  {journal} {\bibinfo  {journal} {Phys. Lett. B}\ }\textbf
  {\bibinfo {volume} {823}},\ \bibinfo {pages} {136752} (\bibinfo {year}
  {2021})},\ \Eprint {http://arxiv.org/abs/2106.04226} {arXiv:2106.04226
  [hep-th]} \BibitemShut {NoStop}%
\bibitem [{\citenamefont {Choi}\ \emph {et~al.}(2021)\citenamefont {Choi},
  \citenamefont {Lin}, \citenamefont {Visinelli},\ and\ \citenamefont
  {Yanagida}}]{Choi:2021aze}%
  \BibitemOpen
  \bibfield  {author} {\bibinfo {author} {\bibfnamefont {G.}~\bibnamefont
  {Choi}}, \bibinfo {author} {\bibfnamefont {W.}~\bibnamefont {Lin}}, \bibinfo
  {author} {\bibfnamefont {L.}~\bibnamefont {Visinelli}}, \ and\ \bibinfo
  {author} {\bibfnamefont {T.~T.}\ \bibnamefont {Yanagida}},\ }\href {\doibase
  10.1103/PhysRevD.104.L101302} {\bibfield  {journal} {\bibinfo  {journal}
  {Phys. Rev. D}\ }\textbf {\bibinfo {volume} {104}},\ \bibinfo {pages}
  {L101302} (\bibinfo {year} {2021})},\ \Eprint
  {http://arxiv.org/abs/2106.12602} {arXiv:2106.12602 [hep-ph]} \BibitemShut
  {NoStop}%
\bibitem [{\citenamefont {Obata}(2021)}]{obata2021implications}%
  \BibitemOpen
  \bibfield  {author} {\bibinfo {author} {\bibfnamefont {I.}~\bibnamefont
  {Obata}},\ }\href@noop {} {\  (\bibinfo {year} {2021})},\ \Eprint
  {http://arxiv.org/abs/2108.02150} {arXiv:2108.02150 [astro-ph.CO]}
  \BibitemShut {NoStop}%
\bibitem [{\citenamefont {Sherwin}\ and\ \citenamefont
  {Namikawa}(2021)}]{sherwin2021cosmic}%
  \BibitemOpen
  \bibfield  {author} {\bibinfo {author} {\bibfnamefont {B.~D.}\ \bibnamefont
  {Sherwin}}\ and\ \bibinfo {author} {\bibfnamefont {T.}~\bibnamefont
  {Namikawa}},\ }\href@noop {} {\  (\bibinfo {year} {2021})},\ \Eprint
  {http://arxiv.org/abs/2108.09287} {arXiv:2108.09287 [astro-ph.CO]}
  \BibitemShut {NoStop}%
\bibitem [{\citenamefont {Mohammadi}\ \emph {et~al.}(2021)\citenamefont
  {Mohammadi}, \citenamefont {Khodagholizadeh}, \citenamefont {Sadegh},\ and\
  \citenamefont {Vahedi}}]{Mohammadi:2021xoh}%
  \BibitemOpen
  \bibfield  {author} {\bibinfo {author} {\bibfnamefont {R.}~\bibnamefont
  {Mohammadi}}, \bibinfo {author} {\bibfnamefont {J.}~\bibnamefont
  {Khodagholizadeh}}, \bibinfo {author} {\bibfnamefont {M.}~\bibnamefont
  {Sadegh}}, \ and\ \bibinfo {author} {\bibfnamefont {A.}~\bibnamefont
  {Vahedi}},\ }\href@noop {} {\  (\bibinfo {year} {2021})},\ \Eprint
  {http://arxiv.org/abs/2109.00152} {arXiv:2109.00152 [hep-ph]} \BibitemShut
  {NoStop}%
\bibitem [{\citenamefont {Diego-Palazuelos}\ \emph {et~al.}(2022)\citenamefont
  {Diego-Palazuelos} \emph {et~al.}}]{diegopalazuelos2022cosmic}%
  \BibitemOpen
  \bibfield  {author} {\bibinfo {author} {\bibfnamefont {P.}~\bibnamefont
  {Diego-Palazuelos}} \emph {et~al.},\ }\href {\doibase
  10.1103/PhysRevLett.128.091302} {\bibfield  {journal} {\bibinfo  {journal}
  {Phys. Rev. Lett.}\ }\textbf {\bibinfo {volume} {128}},\ \bibinfo {pages}
  {091302} (\bibinfo {year} {2022})},\ \Eprint
  {http://arxiv.org/abs/2201.07682} {arXiv:2201.07682 [astro-ph.CO]}
  \BibitemShut {NoStop}%
\bibitem [{\citenamefont {Pogosian}\ \emph {et~al.}(2019)\citenamefont
  {Pogosian}, \citenamefont {Shimon}, \citenamefont {Mewes},\ and\
  \citenamefont {Keating}}]{Pogosian:2019jbt}%
  \BibitemOpen
  \bibfield  {author} {\bibinfo {author} {\bibfnamefont {L.}~\bibnamefont
  {Pogosian}}, \bibinfo {author} {\bibfnamefont {M.}~\bibnamefont {Shimon}},
  \bibinfo {author} {\bibfnamefont {M.}~\bibnamefont {Mewes}}, \ and\ \bibinfo
  {author} {\bibfnamefont {B.}~\bibnamefont {Keating}},\ }\href {\doibase
  10.1103/PhysRevD.100.023507} {\bibfield  {journal} {\bibinfo  {journal}
  {Phys. Rev. D}\ }\textbf {\bibinfo {volume} {100}},\ \bibinfo {pages}
  {023507} (\bibinfo {year} {2019})},\ \Eprint
  {http://arxiv.org/abs/1904.07855} {arXiv:1904.07855 [astro-ph.CO]}
  \BibitemShut {NoStop}%
\bibitem [{\citenamefont {McAllister}\ \emph {et~al.}(2010)\citenamefont
  {McAllister}, \citenamefont {Silverstein},\ and\ \citenamefont
  {Westphal}}]{McAllister2008}%
  \BibitemOpen
  \bibfield  {author} {\bibinfo {author} {\bibfnamefont {L.}~\bibnamefont
  {McAllister}}, \bibinfo {author} {\bibfnamefont {E.}~\bibnamefont
  {Silverstein}}, \ and\ \bibinfo {author} {\bibfnamefont {A.}~\bibnamefont
  {Westphal}},\ }\href {\doibase 10.1103/PhysRevD.82.046003} {\bibfield
  {journal} {\bibinfo  {journal} {Phys. Rev. D}\ }\textbf {\bibinfo {volume}
  {82}},\ \bibinfo {pages} {046003} (\bibinfo {year} {2010})},\ \Eprint
  {http://arxiv.org/abs/0808.0706} {arXiv:0808.0706 [hep-th]} \BibitemShut
  {NoStop}%
\bibitem [{\citenamefont {Silverstein}\ and\ \citenamefont
  {Westphal}(2008)}]{Silverstein}%
  \BibitemOpen
  \bibfield  {author} {\bibinfo {author} {\bibfnamefont {E.}~\bibnamefont
  {Silverstein}}\ and\ \bibinfo {author} {\bibfnamefont {A.}~\bibnamefont
  {Westphal}},\ }\href {\doibase 10.1103/PhysRevD.78.106003} {\bibfield
  {journal} {\bibinfo  {journal} {Phys. Rev. D}\ }\textbf {\bibinfo {volume}
  {78}},\ \bibinfo {pages} {106003} (\bibinfo {year} {2008})},\ \Eprint
  {http://arxiv.org/abs/0803.3085} {arXiv:0803.3085 [hep-th]} \BibitemShut
  {NoStop}%
\bibitem [{\citenamefont {Kaloper}\ and\ \citenamefont
  {Sorbo}(2009{\natexlab{a}})}]{Kaloper:2008fb}%
  \BibitemOpen
  \bibfield  {author} {\bibinfo {author} {\bibfnamefont {N.}~\bibnamefont
  {Kaloper}}\ and\ \bibinfo {author} {\bibfnamefont {L.}~\bibnamefont
  {Sorbo}},\ }\href {\doibase 10.1103/PhysRevLett.102.121301} {\bibfield
  {journal} {\bibinfo  {journal} {Phys. Rev. Lett.}\ }\textbf {\bibinfo
  {volume} {102}},\ \bibinfo {pages} {121301} (\bibinfo {year}
  {2009}{\natexlab{a}})},\ \Eprint {http://arxiv.org/abs/0811.1989}
  {arXiv:0811.1989 [hep-th]} \BibitemShut {NoStop}%
\bibitem [{\citenamefont {Flauger}\ \emph {et~al.}(2010)\citenamefont
  {Flauger}, \citenamefont {McAllister}, \citenamefont {Pajer}, \citenamefont
  {Westphal},\ and\ \citenamefont {Xu}}]{Flauger}%
  \BibitemOpen
  \bibfield  {author} {\bibinfo {author} {\bibfnamefont {R.}~\bibnamefont
  {Flauger}}, \bibinfo {author} {\bibfnamefont {L.}~\bibnamefont {McAllister}},
  \bibinfo {author} {\bibfnamefont {E.}~\bibnamefont {Pajer}}, \bibinfo
  {author} {\bibfnamefont {A.}~\bibnamefont {Westphal}}, \ and\ \bibinfo
  {author} {\bibfnamefont {G.}~\bibnamefont {Xu}},\ }\href {\doibase
  10.1088/1475-7516/2010/06/009} {\bibfield  {journal} {\bibinfo  {journal}
  {JCAP}\ }\textbf {\bibinfo {volume} {06}},\ \bibinfo {pages} {009} (\bibinfo
  {year} {2010})},\ \Eprint {http://arxiv.org/abs/0907.2916} {arXiv:0907.2916
  [hep-th]} \BibitemShut {NoStop}%
\bibitem [{\citenamefont {Kaloper}\ \emph {et~al.}(2011)\citenamefont
  {Kaloper}, \citenamefont {Lawrence},\ and\ \citenamefont
  {Sorbo}}]{Kaloper:2011jz}%
  \BibitemOpen
  \bibfield  {author} {\bibinfo {author} {\bibfnamefont {N.}~\bibnamefont
  {Kaloper}}, \bibinfo {author} {\bibfnamefont {A.}~\bibnamefont {Lawrence}}, \
  and\ \bibinfo {author} {\bibfnamefont {L.}~\bibnamefont {Sorbo}},\ }\href
  {\doibase 10.1088/1475-7516/2011/03/023} {\bibfield  {journal} {\bibinfo
  {journal} {JCAP}\ }\textbf {\bibinfo {volume} {03}},\ \bibinfo {pages} {023}
  (\bibinfo {year} {2011})},\ \Eprint {http://arxiv.org/abs/1101.0026}
  {arXiv:1101.0026 [hep-th]} \BibitemShut {NoStop}%
\bibitem [{\citenamefont {Kaloper}\ and\ \citenamefont
  {Lawrence}(2014)}]{Kaloper:2014zba}%
  \BibitemOpen
  \bibfield  {author} {\bibinfo {author} {\bibfnamefont {N.}~\bibnamefont
  {Kaloper}}\ and\ \bibinfo {author} {\bibfnamefont {A.}~\bibnamefont
  {Lawrence}},\ }\href {\doibase 10.1103/PhysRevD.90.023506} {\bibfield
  {journal} {\bibinfo  {journal} {Phys. Rev. D}\ }\textbf {\bibinfo {volume}
  {90}},\ \bibinfo {pages} {023506} (\bibinfo {year} {2014})},\ \Eprint
  {http://arxiv.org/abs/1404.2912} {arXiv:1404.2912 [hep-th]} \BibitemShut
  {NoStop}%
\bibitem [{\citenamefont {Kaloper}\ and\ \citenamefont
  {Lawrence}(2017)}]{Kaloper:2016fbr}%
  \BibitemOpen
  \bibfield  {author} {\bibinfo {author} {\bibfnamefont {N.}~\bibnamefont
  {Kaloper}}\ and\ \bibinfo {author} {\bibfnamefont {A.}~\bibnamefont
  {Lawrence}},\ }\href {\doibase 10.1103/PhysRevD.95.063526} {\bibfield
  {journal} {\bibinfo  {journal} {Phys. Rev. D}\ }\textbf {\bibinfo {volume}
  {95}},\ \bibinfo {pages} {063526} (\bibinfo {year} {2017})},\ \Eprint
  {http://arxiv.org/abs/1607.06105} {arXiv:1607.06105 [hep-th]} \BibitemShut
  {NoStop}%
\bibitem [{\citenamefont {D'Amico}\ \emph {et~al.}(2021)\citenamefont
  {D'Amico}, \citenamefont {Kaloper},\ and\ \citenamefont
  {Westphal}}]{DAmico:2021vka}%
  \BibitemOpen
  \bibfield  {author} {\bibinfo {author} {\bibfnamefont {G.}~\bibnamefont
  {D'Amico}}, \bibinfo {author} {\bibfnamefont {N.}~\bibnamefont {Kaloper}}, \
  and\ \bibinfo {author} {\bibfnamefont {A.}~\bibnamefont {Westphal}},\ }\href
  {\doibase 10.1103/PhysRevD.104.L081302} {\bibfield  {journal} {\bibinfo
  {journal} {Phys. Rev. D}\ }\textbf {\bibinfo {volume} {104}},\ \bibinfo
  {pages} {L081302} (\bibinfo {year} {2021})},\ \Eprint
  {http://arxiv.org/abs/2101.05861} {arXiv:2101.05861 [hep-th]} \BibitemShut
  {NoStop}%
\bibitem [{\citenamefont {Kaloper}\ and\ \citenamefont
  {Sorbo}(2009{\natexlab{b}})}]{Kaloper:2008qs}%
  \BibitemOpen
  \bibfield  {author} {\bibinfo {author} {\bibfnamefont {N.}~\bibnamefont
  {Kaloper}}\ and\ \bibinfo {author} {\bibfnamefont {L.}~\bibnamefont
  {Sorbo}},\ }\href {\doibase 10.1103/PhysRevD.79.043528} {\bibfield  {journal}
  {\bibinfo  {journal} {Phys. Rev. D}\ }\textbf {\bibinfo {volume} {79}},\
  \bibinfo {pages} {043528} (\bibinfo {year} {2009}{\natexlab{b}})},\ \Eprint
  {http://arxiv.org/abs/0810.5346} {arXiv:0810.5346 [hep-th]} \BibitemShut
  {NoStop}%
\bibitem [{\citenamefont {Panda}\ \emph {et~al.}(2011)\citenamefont {Panda},
  \citenamefont {Sumitomo},\ and\ \citenamefont {Trivedi}}]{Panda:2010uq}%
  \BibitemOpen
  \bibfield  {author} {\bibinfo {author} {\bibfnamefont {S.}~\bibnamefont
  {Panda}}, \bibinfo {author} {\bibfnamefont {Y.}~\bibnamefont {Sumitomo}}, \
  and\ \bibinfo {author} {\bibfnamefont {S.~P.}\ \bibnamefont {Trivedi}},\
  }\href {\doibase 10.1103/PhysRevD.83.083506} {\bibfield  {journal} {\bibinfo
  {journal} {Phys. Rev. D}\ }\textbf {\bibinfo {volume} {83}},\ \bibinfo
  {pages} {083506} (\bibinfo {year} {2011})},\ \Eprint
  {http://arxiv.org/abs/1011.5877} {arXiv:1011.5877 [hep-th]} \BibitemShut
  {NoStop}%
\bibitem [{\citenamefont {D'Amico}\ \emph {et~al.}(2016)\citenamefont
  {D'Amico}, \citenamefont {Hamill},\ and\ \citenamefont
  {Kaloper}}]{DAmico:2016jbm}%
  \BibitemOpen
  \bibfield  {author} {\bibinfo {author} {\bibfnamefont {G.}~\bibnamefont
  {D'Amico}}, \bibinfo {author} {\bibfnamefont {T.}~\bibnamefont {Hamill}}, \
  and\ \bibinfo {author} {\bibfnamefont {N.}~\bibnamefont {Kaloper}},\ }\href
  {\doibase 10.1103/PhysRevD.94.103526} {\bibfield  {journal} {\bibinfo
  {journal} {Phys. Rev. D}\ }\textbf {\bibinfo {volume} {94}},\ \bibinfo
  {pages} {103526} (\bibinfo {year} {2016})},\ \Eprint
  {http://arxiv.org/abs/1605.00996} {arXiv:1605.00996 [hep-th]} \BibitemShut
  {NoStop}%
\bibitem [{\citenamefont {D'Amico}\ \emph {et~al.}(2019)\citenamefont
  {D'Amico}, \citenamefont {Kaloper},\ and\ \citenamefont
  {Lawrence}}]{DAmico:2018mnx}%
  \BibitemOpen
  \bibfield  {author} {\bibinfo {author} {\bibfnamefont {G.}~\bibnamefont
  {D'Amico}}, \bibinfo {author} {\bibfnamefont {N.}~\bibnamefont {Kaloper}}, \
  and\ \bibinfo {author} {\bibfnamefont {A.}~\bibnamefont {Lawrence}},\ }\href
  {\doibase 10.1103/PhysRevD.100.103504} {\bibfield  {journal} {\bibinfo
  {journal} {Phys. Rev. D}\ }\textbf {\bibinfo {volume} {100}},\ \bibinfo
  {pages} {103504} (\bibinfo {year} {2019})},\ \Eprint
  {http://arxiv.org/abs/1809.05109} {arXiv:1809.05109 [hep-th]} \BibitemShut
  {NoStop}%
\bibitem [{\citenamefont {Kaloper}(2019)}]{Kaloper:2018kma}%
  \BibitemOpen
  \bibfield  {author} {\bibinfo {author} {\bibfnamefont {N.}~\bibnamefont
  {Kaloper}},\ }\href {\doibase 10.1007/JHEP11(2019)106} {\bibfield  {journal}
  {\bibinfo  {journal} {JHEP}\ }\textbf {\bibinfo {volume} {11}},\ \bibinfo
  {pages} {106} (\bibinfo {year} {2019})},\ \Eprint
  {http://arxiv.org/abs/1806.03308} {arXiv:1806.03308 [hep-th]} \BibitemShut
  {NoStop}%
\bibitem [{\citenamefont {Kaloper}\ and\ \citenamefont
  {Sorbo}(2006)}]{Kaloper_2006}%
  \BibitemOpen
  \bibfield  {author} {\bibinfo {author} {\bibfnamefont {N.}~\bibnamefont
  {Kaloper}}\ and\ \bibinfo {author} {\bibfnamefont {L.}~\bibnamefont
  {Sorbo}},\ }\href {\doibase 10.1088/1475-7516/2006/04/007} {\bibfield
  {journal} {\bibinfo  {journal} {JCAP}\ }\textbf {\bibinfo {volume} {04}},\
  \bibinfo {pages} {007} (\bibinfo {year} {2006})},\ \Eprint
  {http://arxiv.org/abs/astro-ph/0511543} {arXiv:astro-ph/0511543} \BibitemShut
  {NoStop}%
\bibitem [{\citenamefont {Dutta}\ and\ \citenamefont
  {Scherrer}(2008)}]{Hilltop2008}%
  \BibitemOpen
  \bibfield  {author} {\bibinfo {author} {\bibfnamefont {S.}~\bibnamefont
  {Dutta}}\ and\ \bibinfo {author} {\bibfnamefont {R.~J.}\ \bibnamefont
  {Scherrer}},\ }\href {\doibase 10.1103/PhysRevD.78.123525} {\bibfield
  {journal} {\bibinfo  {journal} {Phys. Rev. D}\ }\textbf {\bibinfo {volume}
  {78}},\ \bibinfo {pages} {123525} (\bibinfo {year} {2008})},\ \Eprint
  {http://arxiv.org/abs/0809.4441} {arXiv:0809.4441 [astro-ph]} \BibitemShut
  {NoStop}%
\bibitem [{\citenamefont {Eskilt}(2022)}]{eskilt2022}%
  \BibitemOpen
  \bibfield  {author} {\bibinfo {author} {\bibfnamefont {J.~R.}\ \bibnamefont
  {Eskilt}},\ }\href@noop {} {\  (\bibinfo {year} {2022})},\ \Eprint
  {http://arxiv.org/abs/2201.13347} {arXiv:2201.13347 [astro-ph.CO]}
  \BibitemShut {NoStop}%
\bibitem [{\citenamefont {Namikawa}\ \emph {et~al.}(2020)\citenamefont
  {Namikawa} \emph {et~al.}}]{Namikawa:2020ffr}%
  \BibitemOpen
  \bibfield  {author} {\bibinfo {author} {\bibfnamefont {T.}~\bibnamefont
  {Namikawa}} \emph {et~al.},\ }\href {\doibase 10.1103/PhysRevD.101.083527}
  {\bibfield  {journal} {\bibinfo  {journal} {Phys. Rev. D}\ }\textbf {\bibinfo
  {volume} {101}},\ \bibinfo {pages} {083527} (\bibinfo {year} {2020})},\
  \Eprint {http://arxiv.org/abs/2001.10465} {arXiv:2001.10465 [astro-ph.CO]}
  \BibitemShut {NoStop}%
\bibitem [{\citenamefont {Bianchini}\ \emph {et~al.}(2020)\citenamefont
  {Bianchini} \emph {et~al.}}]{Bianchini_2020}%
  \BibitemOpen
  \bibfield  {author} {\bibinfo {author} {\bibfnamefont {F.}~\bibnamefont
  {Bianchini}} \emph {et~al.} (\bibinfo {collaboration} {SPT}),\ }\href
  {\doibase 10.1103/PhysRevD.102.083504} {\bibfield  {journal} {\bibinfo
  {journal} {Phys. Rev. D}\ }\textbf {\bibinfo {volume} {102}},\ \bibinfo
  {pages} {083504} (\bibinfo {year} {2020})},\ \Eprint
  {http://arxiv.org/abs/2006.08061} {arXiv:2006.08061 [astro-ph.CO]}
  \BibitemShut {NoStop}%
\bibitem [{\citenamefont {Greco}\ \emph {et~al.}(2022)\citenamefont {Greco},
  \citenamefont {Bartolo},\ and\ \citenamefont {Gruppuso}}]{Greco:2022ufo}%
  \BibitemOpen
  \bibfield  {author} {\bibinfo {author} {\bibfnamefont {A.}~\bibnamefont
  {Greco}}, \bibinfo {author} {\bibfnamefont {N.}~\bibnamefont {Bartolo}}, \
  and\ \bibinfo {author} {\bibfnamefont {A.}~\bibnamefont {Gruppuso}},\ }\href
  {\doibase 10.1088/1475-7516/2022/03/050} {\bibfield  {journal} {\bibinfo
  {journal} {JCAP}\ }\textbf {\bibinfo {volume} {03}},\ \bibinfo {pages} {050}
  (\bibinfo {year} {2022})},\ \Eprint {http://arxiv.org/abs/2202.04584}
  {arXiv:2202.04584 [astro-ph.CO]} \BibitemShut {NoStop}%
\bibitem [{\citenamefont {Zhai}\ \emph {et~al.}(2020)\citenamefont {Zhai},
  \citenamefont {Li}, \citenamefont {Li}, \citenamefont {Li},\ and\
  \citenamefont {Zhang}}]{Zhai:2020vob}%
  \BibitemOpen
  \bibfield  {author} {\bibinfo {author} {\bibfnamefont {H.}~\bibnamefont
  {Zhai}}, \bibinfo {author} {\bibfnamefont {S.-Y.}\ \bibnamefont {Li}},
  \bibinfo {author} {\bibfnamefont {M.}~\bibnamefont {Li}}, \bibinfo {author}
  {\bibfnamefont {H.}~\bibnamefont {Li}}, \ and\ \bibinfo {author}
  {\bibfnamefont {X.}~\bibnamefont {Zhang}},\ }\href {\doibase
  10.1088/1475-7516/2020/12/051} {\bibfield  {journal} {\bibinfo  {journal}
  {JCAP}\ }\textbf {\bibinfo {volume} {12}},\ \bibinfo {pages} {051} (\bibinfo
  {year} {2020})},\ \Eprint {http://arxiv.org/abs/2006.01811} {arXiv:2006.01811
  [astro-ph.CO]} \BibitemShut {NoStop}%
\bibitem [{\citenamefont {Caldwell}\ \emph {et~al.}(2011)\citenamefont
  {Caldwell}, \citenamefont {Gluscevic},\ and\ \citenamefont
  {Kamionkowski}}]{Caldwell:2011pu}%
  \BibitemOpen
  \bibfield  {author} {\bibinfo {author} {\bibfnamefont {R.~R.}\ \bibnamefont
  {Caldwell}}, \bibinfo {author} {\bibfnamefont {V.}~\bibnamefont {Gluscevic}},
  \ and\ \bibinfo {author} {\bibfnamefont {M.}~\bibnamefont {Kamionkowski}},\
  }\href {\doibase 10.1103/PhysRevD.84.043504} {\bibfield  {journal} {\bibinfo
  {journal} {Phys. Rev. D}\ }\textbf {\bibinfo {volume} {84}},\ \bibinfo
  {pages} {043504} (\bibinfo {year} {2011})},\ \Eprint
  {http://arxiv.org/abs/1104.1634} {arXiv:1104.1634 [astro-ph.CO]} \BibitemShut
  {NoStop}%
\bibitem [{\citenamefont {Capparelli}\ \emph {et~al.}(2020)\citenamefont
  {Capparelli}, \citenamefont {Caldwell},\ and\ \citenamefont
  {Melchiorri}}]{Capparelli:2019rtn}%
  \BibitemOpen
  \bibfield  {author} {\bibinfo {author} {\bibfnamefont {L.~M.}\ \bibnamefont
  {Capparelli}}, \bibinfo {author} {\bibfnamefont {R.~R.}\ \bibnamefont
  {Caldwell}}, \ and\ \bibinfo {author} {\bibfnamefont {A.}~\bibnamefont
  {Melchiorri}},\ }\href {\doibase 10.1103/PhysRevD.101.123529} {\bibfield
  {journal} {\bibinfo  {journal} {Phys. Rev. D}\ }\textbf {\bibinfo {volume}
  {101}},\ \bibinfo {pages} {123529} (\bibinfo {year} {2020})},\ \Eprint
  {http://arxiv.org/abs/1909.04621} {arXiv:1909.04621 [astro-ph.CO]}
  \BibitemShut {NoStop}%
\bibitem [{\citenamefont {Agrawal}\ \emph {et~al.}(2020)\citenamefont
  {Agrawal}, \citenamefont {Hook},\ and\ \citenamefont {Huang}}]{Agrawal_2020}%
  \BibitemOpen
  \bibfield  {author} {\bibinfo {author} {\bibfnamefont {P.}~\bibnamefont
  {Agrawal}}, \bibinfo {author} {\bibfnamefont {A.}~\bibnamefont {Hook}}, \
  and\ \bibinfo {author} {\bibfnamefont {J.}~\bibnamefont {Huang}},\ }\href
  {\doibase 10.1007/JHEP07(2020)138} {\bibfield  {journal} {\bibinfo  {journal}
  {JHEP}\ }\textbf {\bibinfo {volume} {07}},\ \bibinfo {pages} {138} (\bibinfo
  {year} {2020})},\ \Eprint {http://arxiv.org/abs/1912.02823} {arXiv:1912.02823
  [astro-ph.CO]} \BibitemShut {NoStop}%
\bibitem [{\citenamefont {Zaldarriaga}\ and\ \citenamefont
  {Seljak}(1997)}]{Zaldarriaga_1997}%
  \BibitemOpen
  \bibfield  {author} {\bibinfo {author} {\bibfnamefont {M.}~\bibnamefont
  {Zaldarriaga}}\ and\ \bibinfo {author} {\bibfnamefont {U.}~\bibnamefont
  {Seljak}},\ }\href {\doibase 10.1103/PhysRevD.55.1830} {\bibfield  {journal}
  {\bibinfo  {journal} {Phys. Rev. D}\ }\textbf {\bibinfo {volume} {55}},\
  \bibinfo {pages} {1830} (\bibinfo {year} {1997})},\ \Eprint
  {http://arxiv.org/abs/astro-ph/9609170} {arXiv:astro-ph/9609170} \BibitemShut
  {NoStop}%
\bibitem [{\citenamefont {Kamionkowski}\ \emph {et~al.}(1997)\citenamefont
  {Kamionkowski}, \citenamefont {Kosowsky},\ and\ \citenamefont
  {Stebbins}}]{Kamionkowski_1997}%
  \BibitemOpen
  \bibfield  {author} {\bibinfo {author} {\bibfnamefont {M.}~\bibnamefont
  {Kamionkowski}}, \bibinfo {author} {\bibfnamefont {A.}~\bibnamefont
  {Kosowsky}}, \ and\ \bibinfo {author} {\bibfnamefont {A.}~\bibnamefont
  {Stebbins}},\ }\href {\doibase 10.1103/PhysRevD.55.7368} {\bibfield
  {journal} {\bibinfo  {journal} {Phys. Rev. D}\ }\textbf {\bibinfo {volume}
  {55}},\ \bibinfo {pages} {7368} (\bibinfo {year} {1997})},\ \Eprint
  {http://arxiv.org/abs/astro-ph/9611125} {arXiv:astro-ph/9611125} \BibitemShut
  {NoStop}%
\bibitem [{\citenamefont {Feng}\ \emph {et~al.}(2005)\citenamefont {Feng},
  \citenamefont {Li}, \citenamefont {Li},\ and\ \citenamefont
  {Zhang}}]{Feng_2005}%
  \BibitemOpen
  \bibfield  {author} {\bibinfo {author} {\bibfnamefont {B.}~\bibnamefont
  {Feng}}, \bibinfo {author} {\bibfnamefont {H.}~\bibnamefont {Li}}, \bibinfo
  {author} {\bibfnamefont {M.-z.}\ \bibnamefont {Li}}, \ and\ \bibinfo {author}
  {\bibfnamefont {X.-m.}\ \bibnamefont {Zhang}},\ }\href {\doibase
  10.1016/j.physletb.2005.06.009} {\bibfield  {journal} {\bibinfo  {journal}
  {Phys. Lett. B}\ }\textbf {\bibinfo {volume} {620}},\ \bibinfo {pages} {27}
  (\bibinfo {year} {2005})},\ \Eprint {http://arxiv.org/abs/hep-ph/0406269}
  {arXiv:hep-ph/0406269} \BibitemShut {NoStop}%
\bibitem [{\citenamefont {Feng}\ \emph {et~al.}(2006)\citenamefont {Feng},
  \citenamefont {Li}, \citenamefont {Xia}, \citenamefont {Chen},\ and\
  \citenamefont {Zhang}}]{Feng:2006dp}%
  \BibitemOpen
  \bibfield  {author} {\bibinfo {author} {\bibfnamefont {B.}~\bibnamefont
  {Feng}}, \bibinfo {author} {\bibfnamefont {M.}~\bibnamefont {Li}}, \bibinfo
  {author} {\bibfnamefont {J.-Q.}\ \bibnamefont {Xia}}, \bibinfo {author}
  {\bibfnamefont {X.}~\bibnamefont {Chen}}, \ and\ \bibinfo {author}
  {\bibfnamefont {X.}~\bibnamefont {Zhang}},\ }\href {\doibase
  10.1103/PhysRevLett.96.221302} {\bibfield  {journal} {\bibinfo  {journal}
  {Phys. Rev. Lett.}\ }\textbf {\bibinfo {volume} {96}},\ \bibinfo {pages}
  {221302} (\bibinfo {year} {2006})},\ \Eprint
  {http://arxiv.org/abs/astro-ph/0601095} {arXiv:astro-ph/0601095} \BibitemShut
  {NoStop}%
\bibitem [{\citenamefont {Liu}\ \emph {et~al.}(2006)\citenamefont {Liu},
  \citenamefont {Lee},\ and\ \citenamefont {Ng}}]{Liu:2006uh}%
  \BibitemOpen
  \bibfield  {author} {\bibinfo {author} {\bibfnamefont {G.-C.}\ \bibnamefont
  {Liu}}, \bibinfo {author} {\bibfnamefont {S.}~\bibnamefont {Lee}}, \ and\
  \bibinfo {author} {\bibfnamefont {K.-W.}\ \bibnamefont {Ng}},\ }\href
  {\doibase 10.1103/PhysRevLett.97.161303} {\bibfield  {journal} {\bibinfo
  {journal} {Phys. Rev. Lett.}\ }\textbf {\bibinfo {volume} {97}},\ \bibinfo
  {pages} {161303} (\bibinfo {year} {2006})},\ \Eprint
  {http://arxiv.org/abs/astro-ph/0606248} {arXiv:astro-ph/0606248} \BibitemShut
  {NoStop}%
\bibitem [{\citenamefont {Finelli}\ and\ \citenamefont
  {Galaverni}(2009)}]{Finelli_2009}%
  \BibitemOpen
  \bibfield  {author} {\bibinfo {author} {\bibfnamefont {F.}~\bibnamefont
  {Finelli}}\ and\ \bibinfo {author} {\bibfnamefont {M.}~\bibnamefont
  {Galaverni}},\ }\href {\doibase 10.1103/PhysRevD.79.063002} {\bibfield
  {journal} {\bibinfo  {journal} {Phys. Rev. D}\ }\textbf {\bibinfo {volume}
  {79}},\ \bibinfo {pages} {063002} (\bibinfo {year} {2009})},\ \Eprint
  {http://arxiv.org/abs/0802.4210} {arXiv:0802.4210 [astro-ph]} \BibitemShut
  {NoStop}%
\bibitem [{\citenamefont {Lee}\ \emph {et~al.}(2015)\citenamefont {Lee},
  \citenamefont {Liu},\ and\ \citenamefont {Ng}}]{Lee:2014rpa}%
  \BibitemOpen
  \bibfield  {author} {\bibinfo {author} {\bibfnamefont {S.}~\bibnamefont
  {Lee}}, \bibinfo {author} {\bibfnamefont {G.-C.}\ \bibnamefont {Liu}}, \ and\
  \bibinfo {author} {\bibfnamefont {K.-W.}\ \bibnamefont {Ng}},\ }\href
  {\doibase 10.1016/j.physletb.2015.05.038} {\bibfield  {journal} {\bibinfo
  {journal} {Phys. Lett. B}\ }\textbf {\bibinfo {volume} {746}},\ \bibinfo
  {pages} {406} (\bibinfo {year} {2015})},\ \Eprint
  {http://arxiv.org/abs/1403.5585} {arXiv:1403.5585 [astro-ph.CO]} \BibitemShut
  {NoStop}%
\bibitem [{\citenamefont {Saito}\ \emph {et~al.}(2007)\citenamefont {Saito},
  \citenamefont {Ichiki},\ and\ \citenamefont {Taruya}}]{Saito:2007kt}%
  \BibitemOpen
  \bibfield  {author} {\bibinfo {author} {\bibfnamefont {S.}~\bibnamefont
  {Saito}}, \bibinfo {author} {\bibfnamefont {K.}~\bibnamefont {Ichiki}}, \
  and\ \bibinfo {author} {\bibfnamefont {A.}~\bibnamefont {Taruya}},\ }\href
  {\doibase 10.1088/1475-7516/2007/09/002} {\bibfield  {journal} {\bibinfo
  {journal} {JCAP}\ }\textbf {\bibinfo {volume} {09}},\ \bibinfo {pages} {002}
  (\bibinfo {year} {2007})},\ \Eprint {http://arxiv.org/abs/0705.3701}
  {arXiv:0705.3701 [astro-ph]} \BibitemShut {NoStop}%
\bibitem [{\citenamefont {Contaldi}\ \emph {et~al.}(2008)\citenamefont
  {Contaldi}, \citenamefont {Magueijo},\ and\ \citenamefont
  {Smolin}}]{Contaldi:2008yz}%
  \BibitemOpen
  \bibfield  {author} {\bibinfo {author} {\bibfnamefont {C.~R.}\ \bibnamefont
  {Contaldi}}, \bibinfo {author} {\bibfnamefont {J.}~\bibnamefont {Magueijo}},
  \ and\ \bibinfo {author} {\bibfnamefont {L.}~\bibnamefont {Smolin}},\ }\href
  {\doibase 10.1103/PhysRevLett.101.141101} {\bibfield  {journal} {\bibinfo
  {journal} {Phys. Rev. Lett.}\ }\textbf {\bibinfo {volume} {101}},\ \bibinfo
  {pages} {141101} (\bibinfo {year} {2008})},\ \Eprint
  {http://arxiv.org/abs/0806.3082} {arXiv:0806.3082 [astro-ph]} \BibitemShut
  {NoStop}%
\bibitem [{\citenamefont {Sorbo}(2011)}]{Sorbo2011}%
  \BibitemOpen
  \bibfield  {author} {\bibinfo {author} {\bibfnamefont {L.}~\bibnamefont
  {Sorbo}},\ }\href {\doibase 10.1088/1475-7516/2011/06/003} {\bibfield
  {journal} {\bibinfo  {journal} {JCAP}\ }\textbf {\bibinfo {volume} {06}},\
  \bibinfo {pages} {003} (\bibinfo {year} {2011})},\ \Eprint
  {http://arxiv.org/abs/1101.1525} {arXiv:1101.1525 [astro-ph.CO]} \BibitemShut
  {NoStop}%
\bibitem [{\citenamefont {Namba}\ \emph {et~al.}(2016)\citenamefont {Namba},
  \citenamefont {Peloso}, \citenamefont {Shiraishi}, \citenamefont {Sorbo},\
  and\ \citenamefont {Unal}}]{Namba:2015gja}%
  \BibitemOpen
  \bibfield  {author} {\bibinfo {author} {\bibfnamefont {R.}~\bibnamefont
  {Namba}}, \bibinfo {author} {\bibfnamefont {M.}~\bibnamefont {Peloso}},
  \bibinfo {author} {\bibfnamefont {M.}~\bibnamefont {Shiraishi}}, \bibinfo
  {author} {\bibfnamefont {L.}~\bibnamefont {Sorbo}}, \ and\ \bibinfo {author}
  {\bibfnamefont {C.}~\bibnamefont {Unal}},\ }\href {\doibase
  10.1088/1475-7516/2016/01/041} {\bibfield  {journal} {\bibinfo  {journal}
  {JCAP}\ }\textbf {\bibinfo {volume} {01}},\ \bibinfo {pages} {041} (\bibinfo
  {year} {2016})},\ \Eprint {http://arxiv.org/abs/1509.07521} {arXiv:1509.07521
  [astro-ph.CO]} \BibitemShut {NoStop}%
\bibitem [{\citenamefont {Thorne}\ \emph {et~al.}(2018)\citenamefont {Thorne},
  \citenamefont {Fujita}, \citenamefont {Hazumi}, \citenamefont {Katayama},
  \citenamefont {Komatsu},\ and\ \citenamefont {Shiraishi}}]{Thorne:2017jft}%
  \BibitemOpen
  \bibfield  {author} {\bibinfo {author} {\bibfnamefont {B.}~\bibnamefont
  {Thorne}}, \bibinfo {author} {\bibfnamefont {T.}~\bibnamefont {Fujita}},
  \bibinfo {author} {\bibfnamefont {M.}~\bibnamefont {Hazumi}}, \bibinfo
  {author} {\bibfnamefont {N.}~\bibnamefont {Katayama}}, \bibinfo {author}
  {\bibfnamefont {E.}~\bibnamefont {Komatsu}}, \ and\ \bibinfo {author}
  {\bibfnamefont {M.}~\bibnamefont {Shiraishi}},\ }\href {\doibase
  10.1103/PhysRevD.97.043506} {\bibfield  {journal} {\bibinfo  {journal} {Phys.
  Rev. D}\ }\textbf {\bibinfo {volume} {97}},\ \bibinfo {pages} {043506}
  (\bibinfo {year} {2018})},\ \Eprint {http://arxiv.org/abs/1707.03240}
  {arXiv:1707.03240 [astro-ph.CO]} \BibitemShut {NoStop}%
\bibitem [{\citenamefont {Li}\ and\ \citenamefont {Zhang}(2008)}]{Li_2008}%
  \BibitemOpen
  \bibfield  {author} {\bibinfo {author} {\bibfnamefont {M.}~\bibnamefont
  {Li}}\ and\ \bibinfo {author} {\bibfnamefont {X.}~\bibnamefont {Zhang}},\
  }\href {\doibase 10.1103/PhysRevD.78.103516} {\bibfield  {journal} {\bibinfo
  {journal} {Phys. Rev. D}\ }\textbf {\bibinfo {volume} {78}},\ \bibinfo
  {pages} {103516} (\bibinfo {year} {2008})},\ \Eprint
  {http://arxiv.org/abs/0810.0403} {arXiv:0810.0403 [astro-ph]} \BibitemShut
  {NoStop}%
\bibitem [{\citenamefont {Fedderke}\ \emph {et~al.}(2019)\citenamefont
  {Fedderke}, \citenamefont {Graham},\ and\ \citenamefont
  {Rajendran}}]{Fedderke:2019ajk}%
  \BibitemOpen
  \bibfield  {author} {\bibinfo {author} {\bibfnamefont {M.~A.}\ \bibnamefont
  {Fedderke}}, \bibinfo {author} {\bibfnamefont {P.~W.}\ \bibnamefont
  {Graham}}, \ and\ \bibinfo {author} {\bibfnamefont {S.}~\bibnamefont
  {Rajendran}},\ }\href {\doibase 10.1103/PhysRevD.100.015040} {\bibfield
  {journal} {\bibinfo  {journal} {Phys. Rev. D}\ }\textbf {\bibinfo {volume}
  {100}},\ \bibinfo {pages} {015040} (\bibinfo {year} {2019})},\ \Eprint
  {http://arxiv.org/abs/1903.02666} {arXiv:1903.02666 [astro-ph.CO]}
  \BibitemShut {NoStop}%
\bibitem [{\citenamefont {Hlozek}\ \emph {et~al.}(2015)\citenamefont {Hlozek},
  \citenamefont {Grin}, \citenamefont {Marsh},\ and\ \citenamefont
  {Ferreira}}]{Hlozek:2014lca}%
  \BibitemOpen
  \bibfield  {author} {\bibinfo {author} {\bibfnamefont {R.}~\bibnamefont
  {Hlozek}}, \bibinfo {author} {\bibfnamefont {D.}~\bibnamefont {Grin}},
  \bibinfo {author} {\bibfnamefont {D.~J.~E.}\ \bibnamefont {Marsh}}, \ and\
  \bibinfo {author} {\bibfnamefont {P.~G.}\ \bibnamefont {Ferreira}},\ }\href
  {\doibase 10.1103/PhysRevD.91.103512} {\bibfield  {journal} {\bibinfo
  {journal} {Phys. Rev. D}\ }\textbf {\bibinfo {volume} {91}},\ \bibinfo
  {pages} {103512} (\bibinfo {year} {2015})},\ \Eprint
  {http://arxiv.org/abs/1410.2896} {arXiv:1410.2896 [astro-ph.CO]} \BibitemShut
  {NoStop}%
\bibitem [{\citenamefont {Berg}\ \emph {et~al.}(2017)\citenamefont {Berg},
  \citenamefont {Conlon}, \citenamefont {Day}, \citenamefont {Jennings},
  \citenamefont {Krippendorf}, \citenamefont {Powell},\ and\ \citenamefont
  {Rummel}}]{Berg:2016ese}%
  \BibitemOpen
  \bibfield  {author} {\bibinfo {author} {\bibfnamefont {M.}~\bibnamefont
  {Berg}}, \bibinfo {author} {\bibfnamefont {J.~P.}\ \bibnamefont {Conlon}},
  \bibinfo {author} {\bibfnamefont {F.}~\bibnamefont {Day}}, \bibinfo {author}
  {\bibfnamefont {N.}~\bibnamefont {Jennings}}, \bibinfo {author}
  {\bibfnamefont {S.}~\bibnamefont {Krippendorf}}, \bibinfo {author}
  {\bibfnamefont {A.~J.}\ \bibnamefont {Powell}}, \ and\ \bibinfo {author}
  {\bibfnamefont {M.}~\bibnamefont {Rummel}},\ }\href {\doibase
  10.3847/1538-4357/aa8b16} {\bibfield  {journal} {\bibinfo  {journal}
  {Astrophys. J.}\ }\textbf {\bibinfo {volume} {847}},\ \bibinfo {pages} {101}
  (\bibinfo {year} {2017})},\ \Eprint {http://arxiv.org/abs/1605.01043}
  {arXiv:1605.01043 [astro-ph.HE]} \BibitemShut {NoStop}%
\bibitem [{\citenamefont {Banerjee}\ \emph {et~al.}(2021)\citenamefont
  {Banerjee}, \citenamefont {Cai}, \citenamefont {Heisenberg}, \citenamefont
  {Colg\'ain}, \citenamefont {Sheikh-Jabbari},\ and\ \citenamefont
  {Yang}}]{Banerjee:2020xcn}%
  \BibitemOpen
  \bibfield  {author} {\bibinfo {author} {\bibfnamefont {A.}~\bibnamefont
  {Banerjee}}, \bibinfo {author} {\bibfnamefont {H.}~\bibnamefont {Cai}},
  \bibinfo {author} {\bibfnamefont {L.}~\bibnamefont {Heisenberg}}, \bibinfo
  {author} {\bibfnamefont {E.~O.}\ \bibnamefont {Colg\'ain}}, \bibinfo {author}
  {\bibfnamefont {M.~M.}\ \bibnamefont {Sheikh-Jabbari}}, \ and\ \bibinfo
  {author} {\bibfnamefont {T.}~\bibnamefont {Yang}},\ }\href {\doibase
  10.1103/PhysRevD.103.L081305} {\bibfield  {journal} {\bibinfo  {journal}
  {Phys. Rev. D}\ }\textbf {\bibinfo {volume} {103}},\ \bibinfo {pages}
  {L081305} (\bibinfo {year} {2021})},\ \Eprint
  {http://arxiv.org/abs/2006.00244} {arXiv:2006.00244 [astro-ph.CO]}
  \BibitemShut {NoStop}%
\bibitem [{\citenamefont {Heisenberg}\ \emph
  {et~al.}(2022{\natexlab{a}})\citenamefont {Heisenberg}, \citenamefont
  {Villarrubia-Rojo},\ and\ \citenamefont {Zosso}}]{Heisenberg:2022gqk}%
  \BibitemOpen
  \bibfield  {author} {\bibinfo {author} {\bibfnamefont {L.}~\bibnamefont
  {Heisenberg}}, \bibinfo {author} {\bibfnamefont {H.}~\bibnamefont
  {Villarrubia-Rojo}}, \ and\ \bibinfo {author} {\bibfnamefont
  {J.}~\bibnamefont {Zosso}},\ }\href@noop {} {\  (\bibinfo {year}
  {2022}{\natexlab{a}})},\ \Eprint {http://arxiv.org/abs/2202.01202}
  {arXiv:2202.01202 [astro-ph.CO]} \BibitemShut {NoStop}%
\bibitem [{\citenamefont {Heisenberg}\ \emph
  {et~al.}(2022{\natexlab{b}})\citenamefont {Heisenberg}, \citenamefont
  {Villarrubia-Rojo},\ and\ \citenamefont {Zosso}}]{Heisenberg:2022lob}%
  \BibitemOpen
  \bibfield  {author} {\bibinfo {author} {\bibfnamefont {L.}~\bibnamefont
  {Heisenberg}}, \bibinfo {author} {\bibfnamefont {H.}~\bibnamefont
  {Villarrubia-Rojo}}, \ and\ \bibinfo {author} {\bibfnamefont
  {J.}~\bibnamefont {Zosso}},\ }\href@noop {} {\  (\bibinfo {year}
  {2022}{\natexlab{b}})},\ \Eprint {http://arxiv.org/abs/2201.11623}
  {arXiv:2201.11623 [astro-ph.CO]} \BibitemShut {NoStop}%
\bibitem [{\citenamefont {Banks}\ \emph {et~al.}(2003)\citenamefont {Banks},
  \citenamefont {Dine}, \citenamefont {Fox},\ and\ \citenamefont
  {Gorbatov}}]{Banks:2003sx}%
  \BibitemOpen
  \bibfield  {author} {\bibinfo {author} {\bibfnamefont {T.}~\bibnamefont
  {Banks}}, \bibinfo {author} {\bibfnamefont {M.}~\bibnamefont {Dine}},
  \bibinfo {author} {\bibfnamefont {P.~J.}\ \bibnamefont {Fox}}, \ and\
  \bibinfo {author} {\bibfnamefont {E.}~\bibnamefont {Gorbatov}},\ }\href
  {\doibase 10.1088/1475-7516/2003/06/001} {\bibfield  {journal} {\bibinfo
  {journal} {JCAP}\ }\textbf {\bibinfo {volume} {06}},\ \bibinfo {pages} {001}
  (\bibinfo {year} {2003})},\ \Eprint {http://arxiv.org/abs/hep-th/0303252}
  {arXiv:hep-th/0303252} \BibitemShut {NoStop}%
\bibitem [{\citenamefont {Arkani-Hamed}\ \emph {et~al.}(2007)\citenamefont
  {Arkani-Hamed}, \citenamefont {Motl}, \citenamefont {Nicolis},\ and\
  \citenamefont {Vafa}}]{Arkani_Hamed_2007}%
  \BibitemOpen
  \bibfield  {author} {\bibinfo {author} {\bibfnamefont {N.}~\bibnamefont
  {Arkani-Hamed}}, \bibinfo {author} {\bibfnamefont {L.}~\bibnamefont {Motl}},
  \bibinfo {author} {\bibfnamefont {A.}~\bibnamefont {Nicolis}}, \ and\
  \bibinfo {author} {\bibfnamefont {C.}~\bibnamefont {Vafa}},\ }\href {\doibase
  10.1088/1126-6708/2007/06/060} {\bibfield  {journal} {\bibinfo  {journal}
  {JHEP}\ }\textbf {\bibinfo {volume} {06}},\ \bibinfo {pages} {060} (\bibinfo
  {year} {2007})},\ \Eprint {http://arxiv.org/abs/hep-th/0601001}
  {arXiv:hep-th/0601001} \BibitemShut {NoStop}%
\bibitem [{\citenamefont {Ibe}\ \emph {et~al.}(2019)\citenamefont {Ibe},
  \citenamefont {Yamazaki},\ and\ \citenamefont {Yanagida}}]{Ibe_2019}%
  \BibitemOpen
  \bibfield  {author} {\bibinfo {author} {\bibfnamefont {M.}~\bibnamefont
  {Ibe}}, \bibinfo {author} {\bibfnamefont {M.}~\bibnamefont {Yamazaki}}, \
  and\ \bibinfo {author} {\bibfnamefont {T.~T.}\ \bibnamefont {Yanagida}},\
  }\href {\doibase 10.1088/1361-6382/ab5197} {\bibfield  {journal} {\bibinfo
  {journal} {Class. Quant. Grav.}\ }\textbf {\bibinfo {volume} {36}},\ \bibinfo
  {pages} {235020} (\bibinfo {year} {2019})},\ \Eprint
  {http://arxiv.org/abs/1811.04664} {arXiv:1811.04664 [hep-th]} \BibitemShut
  {NoStop}%
\bibitem [{\citenamefont {Cicoli}\ \emph {et~al.}(2021)\citenamefont {Cicoli},
  \citenamefont {Cunillera}, \citenamefont {Padilla},\ and\ \citenamefont
  {Pedro}}]{Cicoli:2021skd}%
  \BibitemOpen
  \bibfield  {author} {\bibinfo {author} {\bibfnamefont {M.}~\bibnamefont
  {Cicoli}}, \bibinfo {author} {\bibfnamefont {F.}~\bibnamefont {Cunillera}},
  \bibinfo {author} {\bibfnamefont {A.}~\bibnamefont {Padilla}}, \ and\
  \bibinfo {author} {\bibfnamefont {F.~G.}\ \bibnamefont {Pedro}},\ }\href@noop
  {} {\  (\bibinfo {year} {2021})},\ \Eprint {http://arxiv.org/abs/2112.10783}
  {arXiv:2112.10783 [hep-th]} \BibitemShut {NoStop}%
\bibitem [{\citenamefont {Reig}(2021)}]{stochastic2021}%
  \BibitemOpen
  \bibfield  {author} {\bibinfo {author} {\bibfnamefont {M.}~\bibnamefont
  {Reig}},\ }\href {\doibase 10.1007/JHEP09(2021)207} {\bibfield  {journal}
  {\bibinfo  {journal} {JHEP}\ }\textbf {\bibinfo {volume} {09}},\ \bibinfo
  {pages} {207} (\bibinfo {year} {2021})},\ \Eprint
  {http://arxiv.org/abs/2104.09923} {arXiv:2104.09923 [hep-ph]} \BibitemShut
  {NoStop}%
\bibitem [{\citenamefont {Kamionkowski}\ \emph {et~al.}(2014)\citenamefont
  {Kamionkowski}, \citenamefont {Pradler},\ and\ \citenamefont
  {Walker}}]{Kamionkowski_2014}%
  \BibitemOpen
  \bibfield  {author} {\bibinfo {author} {\bibfnamefont {M.}~\bibnamefont
  {Kamionkowski}}, \bibinfo {author} {\bibfnamefont {J.}~\bibnamefont
  {Pradler}}, \ and\ \bibinfo {author} {\bibfnamefont {D.~G.~E.}\ \bibnamefont
  {Walker}},\ }\href {\doibase 10.1103/PhysRevLett.113.251302} {\bibfield
  {journal} {\bibinfo  {journal} {Phys. Rev. Lett.}\ }\textbf {\bibinfo
  {volume} {113}},\ \bibinfo {pages} {251302} (\bibinfo {year} {2014})},\
  \Eprint {http://arxiv.org/abs/1409.0549} {arXiv:1409.0549 [hep-ph]}
  \BibitemShut {NoStop}%
\bibitem [{\citenamefont {McAllister}\ \emph {et~al.}(2014)\citenamefont
  {McAllister}, \citenamefont {Silverstein}, \citenamefont {Westphal},\ and\
  \citenamefont {Wrase}}]{monodromypower}%
  \BibitemOpen
  \bibfield  {author} {\bibinfo {author} {\bibfnamefont {L.}~\bibnamefont
  {McAllister}}, \bibinfo {author} {\bibfnamefont {E.}~\bibnamefont
  {Silverstein}}, \bibinfo {author} {\bibfnamefont {A.}~\bibnamefont
  {Westphal}}, \ and\ \bibinfo {author} {\bibfnamefont {T.}~\bibnamefont
  {Wrase}},\ }\href {\doibase 10.1007/JHEP09(2014)123} {\bibfield  {journal}
  {\bibinfo  {journal} {JHEP}\ }\textbf {\bibinfo {volume} {09}},\ \bibinfo
  {pages} {123} (\bibinfo {year} {2014})},\ \Eprint
  {http://arxiv.org/abs/1405.3652} {arXiv:1405.3652 [hep-th]} \BibitemShut
  {NoStop}%
\bibitem [{\citenamefont {Gupta}\ \emph {et~al.}(2012)\citenamefont {Gupta},
  \citenamefont {Panda},\ and\ \citenamefont {Sen}}]{Gupta2012}%
  \BibitemOpen
  \bibfield  {author} {\bibinfo {author} {\bibfnamefont {G.}~\bibnamefont
  {Gupta}}, \bibinfo {author} {\bibfnamefont {S.}~\bibnamefont {Panda}}, \ and\
  \bibinfo {author} {\bibfnamefont {A.~A.}\ \bibnamefont {Sen}},\ }\href
  {\doibase 10.1103/PhysRevD.85.089907} {\bibfield  {journal} {\bibinfo
  {journal} {Phys. Rev. D}\ }\textbf {\bibinfo {volume} {85}},\ \bibinfo
  {pages} {023501} (\bibinfo {year} {2012})},\ \Eprint
  {http://arxiv.org/abs/1108.1322} {arXiv:1108.1322 [astro-ph.CO]} \BibitemShut
  {NoStop}%
\bibitem [{\citenamefont {Aghanim}\ \emph {et~al.}(2020)\citenamefont {Aghanim}
  \emph {et~al.}}]{Planck:2018vyg}%
  \BibitemOpen
  \bibfield  {author} {\bibinfo {author} {\bibfnamefont {N.}~\bibnamefont
  {Aghanim}} \emph {et~al.} (\bibinfo {collaboration} {Planck}),\ }\href
  {\doibase 10.1051/0004-6361/201833910} {\bibfield  {journal} {\bibinfo
  {journal} {Astron. Astrophys.}\ }\textbf {\bibinfo {volume} {641}},\ \bibinfo
  {pages} {A6} (\bibinfo {year} {2020})},\ \bibinfo {note} {[Erratum:
  Astron.Astrophys. 652, C4 (2021)]},\ \Eprint
  {http://arxiv.org/abs/1807.06209} {arXiv:1807.06209 [astro-ph.CO]}
  \BibitemShut {NoStop}%
\bibitem [{\citenamefont {Weinberg}(2008)}]{Weinbergcos}%
  \BibitemOpen
  \bibfield  {author} {\bibinfo {author} {\bibfnamefont {S.}~\bibnamefont
  {Weinberg}},\ }\href@noop {} {\emph {\bibinfo {title} {{Cosmology}}}}\
  (\bibinfo {year} {2008})\BibitemShut {NoStop}%
\bibitem [{\citenamefont {Reyn\'es}\ \emph {et~al.}(2021)\citenamefont
  {Reyn\'es}, \citenamefont {Matthews}, \citenamefont {Reynolds}, \citenamefont
  {Russell}, \citenamefont {Smith},\ and\ \citenamefont
  {Marsh}}]{reynes2021new}%
  \BibitemOpen
  \bibfield  {author} {\bibinfo {author} {\bibfnamefont {J.~S.}\ \bibnamefont
  {Reyn\'es}}, \bibinfo {author} {\bibfnamefont {J.~H.}\ \bibnamefont
  {Matthews}}, \bibinfo {author} {\bibfnamefont {C.~S.}\ \bibnamefont
  {Reynolds}}, \bibinfo {author} {\bibfnamefont {H.~R.}\ \bibnamefont
  {Russell}}, \bibinfo {author} {\bibfnamefont {R.~N.}\ \bibnamefont {Smith}},
  \ and\ \bibinfo {author} {\bibfnamefont {M.~C.~D.}\ \bibnamefont {Marsh}},\
  }\href {\doibase 10.1093/mnras/stab3464} {\bibfield  {journal} {\bibinfo
  {journal} {Mon. Not. Roy. Astron. Soc.}\ }\textbf {\bibinfo {volume} {510}},\
  \bibinfo {pages} {1264} (\bibinfo {year} {2021})},\ \Eprint
  {http://arxiv.org/abs/2109.03261} {arXiv:2109.03261 [astro-ph.HE]}
  \BibitemShut {NoStop}%
\bibitem [{\citenamefont {Caldwell}\ and\ \citenamefont
  {Linder}(2005)}]{Caldwell2005}%
  \BibitemOpen
  \bibfield  {author} {\bibinfo {author} {\bibfnamefont {R.~R.}\ \bibnamefont
  {Caldwell}}\ and\ \bibinfo {author} {\bibfnamefont {E.~V.}\ \bibnamefont
  {Linder}},\ }\href {\doibase 10.1103/PhysRevLett.95.141301} {\bibfield
  {journal} {\bibinfo  {journal} {Phys. Rev. Lett.}\ }\textbf {\bibinfo
  {volume} {95}},\ \bibinfo {pages} {141301} (\bibinfo {year} {2005})},\
  \Eprint {http://arxiv.org/abs/astro-ph/0505494} {arXiv:astro-ph/0505494}
  \BibitemShut {NoStop}%
\bibitem [{\citenamefont {Scherrer}\ and\ \citenamefont
  {Sen}(2008)}]{thawingquintessence}%
  \BibitemOpen
  \bibfield  {author} {\bibinfo {author} {\bibfnamefont {R.~J.}\ \bibnamefont
  {Scherrer}}\ and\ \bibinfo {author} {\bibfnamefont {A.~A.}\ \bibnamefont
  {Sen}},\ }\href {\doibase 10.1103/PhysRevD.77.083515} {\bibfield  {journal}
  {\bibinfo  {journal} {Phys. Rev. D}\ }\textbf {\bibinfo {volume} {77}},\
  \bibinfo {pages} {083515} (\bibinfo {year} {2008})},\ \Eprint
  {http://arxiv.org/abs/0712.3450} {arXiv:0712.3450 [astro-ph]} \BibitemShut
  {NoStop}%
\bibitem [{\citenamefont {Linder}(2015)}]{quintlaststand}%
  \BibitemOpen
  \bibfield  {author} {\bibinfo {author} {\bibfnamefont {E.~V.}\ \bibnamefont
  {Linder}},\ }\href {\doibase 10.1103/PhysRevD.91.063006} {\bibfield
  {journal} {\bibinfo  {journal} {Phys. Rev. D}\ }\textbf {\bibinfo {volume}
  {91}},\ \bibinfo {pages} {063006} (\bibinfo {year} {2015})},\ \Eprint
  {http://arxiv.org/abs/1501.01634} {arXiv:1501.01634 [astro-ph.CO]}
  \BibitemShut {NoStop}%
\bibitem [{\citenamefont {Capone}\ \emph {et~al.}(2006)\citenamefont {Capone},
  \citenamefont {Rubano},\ and\ \citenamefont {Scudellaro}}]{SR2006}%
  \BibitemOpen
  \bibfield  {author} {\bibinfo {author} {\bibfnamefont {M.}~\bibnamefont
  {Capone}}, \bibinfo {author} {\bibfnamefont {C.}~\bibnamefont {Rubano}}, \
  and\ \bibinfo {author} {\bibfnamefont {P.}~\bibnamefont {Scudellaro}},\
  }\href {\doibase 10.1209/epl/i2005-10350-5} {\bibfield  {journal} {\bibinfo
  {journal} {Europhys. Lett.}\ }\textbf {\bibinfo {volume} {73}},\ \bibinfo
  {pages} {149} (\bibinfo {year} {2006})},\ \Eprint
  {http://arxiv.org/abs/astro-ph/0607556} {arXiv:astro-ph/0607556} \BibitemShut
  {NoStop}%
\bibitem [{\citenamefont {Linder}(2020)}]{PoleDE}%
  \BibitemOpen
  \bibfield  {author} {\bibinfo {author} {\bibfnamefont {E.~V.}\ \bibnamefont
  {Linder}},\ }\href {\doibase 10.1103/PhysRevD.101.023506} {\bibfield
  {journal} {\bibinfo  {journal} {Phys. Rev. D}\ }\textbf {\bibinfo {volume}
  {101}},\ \bibinfo {pages} {023506} (\bibinfo {year} {2020})},\ \Eprint
  {http://arxiv.org/abs/1911.01606} {arXiv:1911.01606 [astro-ph.CO]}
  \BibitemShut {NoStop}%
\bibitem [{\citenamefont {Marsh}\ \emph {et~al.}(2014)\citenamefont {Marsh},
  \citenamefont {Bull}, \citenamefont {Ferreira},\ and\ \citenamefont
  {Pontzen}}]{Marsh2014}%
  \BibitemOpen
  \bibfield  {author} {\bibinfo {author} {\bibfnamefont {D.~J.~E.}\
  \bibnamefont {Marsh}}, \bibinfo {author} {\bibfnamefont {P.}~\bibnamefont
  {Bull}}, \bibinfo {author} {\bibfnamefont {P.~G.}\ \bibnamefont {Ferreira}},
  \ and\ \bibinfo {author} {\bibfnamefont {A.}~\bibnamefont {Pontzen}},\ }\href
  {\doibase 10.1103/PhysRevD.90.105023} {\bibfield  {journal} {\bibinfo
  {journal} {Phys. Rev. D}\ }\textbf {\bibinfo {volume} {90}},\ \bibinfo
  {pages} {105023} (\bibinfo {year} {2014})},\ \Eprint
  {http://arxiv.org/abs/1406.2301} {arXiv:1406.2301 [astro-ph.CO]} \BibitemShut
  {NoStop}%
\bibitem [{\citenamefont {Dvali}\ and\ \citenamefont
  {Gomez}(2016)}]{Dvali:2014gua}%
  \BibitemOpen
  \bibfield  {author} {\bibinfo {author} {\bibfnamefont {G.}~\bibnamefont
  {Dvali}}\ and\ \bibinfo {author} {\bibfnamefont {C.}~\bibnamefont {Gomez}},\
  }\href {\doibase 10.1002/andp.201500216} {\bibfield  {journal} {\bibinfo
  {journal} {Annalen Phys.}\ }\textbf {\bibinfo {volume} {528}},\ \bibinfo
  {pages} {68} (\bibinfo {year} {2016})},\ \Eprint
  {http://arxiv.org/abs/1412.8077} {arXiv:1412.8077 [hep-th]} \BibitemShut
  {NoStop}%
\bibitem [{\citenamefont {Dvali}\ \emph {et~al.}(2017)\citenamefont {Dvali},
  \citenamefont {Gomez},\ and\ \citenamefont {Zell}}]{Dvali:2017eba}%
  \BibitemOpen
  \bibfield  {author} {\bibinfo {author} {\bibfnamefont {G.}~\bibnamefont
  {Dvali}}, \bibinfo {author} {\bibfnamefont {C.}~\bibnamefont {Gomez}}, \ and\
  \bibinfo {author} {\bibfnamefont {S.}~\bibnamefont {Zell}},\ }\href {\doibase
  10.1088/1475-7516/2017/06/028} {\bibfield  {journal} {\bibinfo  {journal}
  {JCAP}\ }\textbf {\bibinfo {volume} {06}},\ \bibinfo {pages} {028} (\bibinfo
  {year} {2017})},\ \Eprint {http://arxiv.org/abs/1701.08776} {arXiv:1701.08776
  [hep-th]} \BibitemShut {NoStop}%
\bibitem [{\citenamefont {Danielsson}\ and\ \citenamefont
  {Van~Riet}(2018)}]{Danielsson:2018ztv}%
  \BibitemOpen
  \bibfield  {author} {\bibinfo {author} {\bibfnamefont {U.~H.}\ \bibnamefont
  {Danielsson}}\ and\ \bibinfo {author} {\bibfnamefont {T.}~\bibnamefont
  {Van~Riet}},\ }\href {\doibase 10.1142/S0218271818300070} {\bibfield
  {journal} {\bibinfo  {journal} {Int. J. Mod. Phys. D}\ }\textbf {\bibinfo
  {volume} {27}},\ \bibinfo {pages} {1830007} (\bibinfo {year} {2018})},\
  \Eprint {http://arxiv.org/abs/1804.01120} {arXiv:1804.01120 [hep-th]}
  \BibitemShut {NoStop}%
\bibitem [{\citenamefont {Agrawal}\ \emph {et~al.}(2018)\citenamefont
  {Agrawal}, \citenamefont {Obied}, \citenamefont {Steinhardt},\ and\
  \citenamefont {Vafa}}]{Swampland2018}%
  \BibitemOpen
  \bibfield  {author} {\bibinfo {author} {\bibfnamefont {P.}~\bibnamefont
  {Agrawal}}, \bibinfo {author} {\bibfnamefont {G.}~\bibnamefont {Obied}},
  \bibinfo {author} {\bibfnamefont {P.~J.}\ \bibnamefont {Steinhardt}}, \ and\
  \bibinfo {author} {\bibfnamefont {C.}~\bibnamefont {Vafa}},\ }\href {\doibase
  10.1016/j.physletb.2018.07.040} {\bibfield  {journal} {\bibinfo  {journal}
  {Phys. Lett. B}\ }\textbf {\bibinfo {volume} {784}},\ \bibinfo {pages} {271}
  (\bibinfo {year} {2018})},\ \Eprint {http://arxiv.org/abs/1806.09718}
  {arXiv:1806.09718 [hep-th]} \BibitemShut {NoStop}%
\bibitem [{\citenamefont {Obied}\ \emph {et~al.}(2018)\citenamefont {Obied},
  \citenamefont {Ooguri}, \citenamefont {Spodyneiko},\ and\ \citenamefont
  {Vafa}}]{Obied:2018sgi}%
  \BibitemOpen
  \bibfield  {author} {\bibinfo {author} {\bibfnamefont {G.}~\bibnamefont
  {Obied}}, \bibinfo {author} {\bibfnamefont {H.}~\bibnamefont {Ooguri}},
  \bibinfo {author} {\bibfnamefont {L.}~\bibnamefont {Spodyneiko}}, \ and\
  \bibinfo {author} {\bibfnamefont {C.}~\bibnamefont {Vafa}},\ }\href@noop {}
  {\  (\bibinfo {year} {2018})},\ \Eprint {http://arxiv.org/abs/1806.08362}
  {arXiv:1806.08362 [hep-th]} \BibitemShut {NoStop}%
\bibitem [{\citenamefont {Garg}\ and\ \citenamefont
  {Krishnan}(2019)}]{Garg:2018reu}%
  \BibitemOpen
  \bibfield  {author} {\bibinfo {author} {\bibfnamefont {S.~K.}\ \bibnamefont
  {Garg}}\ and\ \bibinfo {author} {\bibfnamefont {C.}~\bibnamefont
  {Krishnan}},\ }\href {\doibase 10.1007/JHEP11(2019)075} {\bibfield  {journal}
  {\bibinfo  {journal} {JHEP}\ }\textbf {\bibinfo {volume} {11}},\ \bibinfo
  {pages} {075} (\bibinfo {year} {2019})},\ \Eprint
  {http://arxiv.org/abs/1807.05193} {arXiv:1807.05193 [hep-th]} \BibitemShut
  {NoStop}%
\bibitem [{\citenamefont {Ooguri}\ \emph {et~al.}(2019)\citenamefont {Ooguri},
  \citenamefont {Palti}, \citenamefont {Shiu},\ and\ \citenamefont
  {Vafa}}]{Ooguri:2018wrx}%
  \BibitemOpen
  \bibfield  {author} {\bibinfo {author} {\bibfnamefont {H.}~\bibnamefont
  {Ooguri}}, \bibinfo {author} {\bibfnamefont {E.}~\bibnamefont {Palti}},
  \bibinfo {author} {\bibfnamefont {G.}~\bibnamefont {Shiu}}, \ and\ \bibinfo
  {author} {\bibfnamefont {C.}~\bibnamefont {Vafa}},\ }\href {\doibase
  10.1016/j.physletb.2018.11.018} {\bibfield  {journal} {\bibinfo  {journal}
  {Phys. Lett. B}\ }\textbf {\bibinfo {volume} {788}},\ \bibinfo {pages} {180}
  (\bibinfo {year} {2019})},\ \Eprint {http://arxiv.org/abs/1810.05506}
  {arXiv:1810.05506 [hep-th]} \BibitemShut {NoStop}%
\bibitem [{\citenamefont {Schmidt}(2017)}]{schmidt2017monodromic}%
  \BibitemOpen
  \bibfield  {author} {\bibinfo {author} {\bibfnamefont {F.}~\bibnamefont
  {Schmidt}},\ }\href@noop {} {\  (\bibinfo {year} {2017})},\ \Eprint
  {http://arxiv.org/abs/1709.01544} {arXiv:1709.01544 [astro-ph.CO]}
  \BibitemShut {NoStop}%
\bibitem [{\citenamefont {{Hazumi}}\ \emph {et~al.}(2019)\citenamefont
  {{Hazumi}} \emph {et~al.}}]{2019JLTP..194..443H}%
  \BibitemOpen
  \bibfield  {author} {\bibinfo {author} {\bibfnamefont {M.}~\bibnamefont
  {{Hazumi}}} \emph {et~al.},\ }\href {\doibase 10.1007/s10909-019-02150-5}
  {\bibfield  {journal} {\bibinfo  {journal} {Journal of Low Temperature
  Physics}\ }\textbf {\bibinfo {volume} {194}},\ \bibinfo {pages} {443}
  (\bibinfo {year} {2019})}\BibitemShut {NoStop}%
\bibitem [{\citenamefont {Nakatsuka}\ \emph {et~al.}(2022)\citenamefont
  {Nakatsuka}, \citenamefont {Namikawa},\ and\ \citenamefont
  {Komatsu}}]{Nakatsuka:2022epj}%
  \BibitemOpen
  \bibfield  {author} {\bibinfo {author} {\bibfnamefont {H.}~\bibnamefont
  {Nakatsuka}}, \bibinfo {author} {\bibfnamefont {T.}~\bibnamefont {Namikawa}},
  \ and\ \bibinfo {author} {\bibfnamefont {E.}~\bibnamefont {Komatsu}},\
  }\href@noop {} {\  (\bibinfo {year} {2022})},\ \Eprint
  {http://arxiv.org/abs/2203.08560} {arXiv:2203.08560 [astro-ph.CO]}
  \BibitemShut {NoStop}%
\bibitem [{\citenamefont {Kumar}\ \emph {et~al.}(2013)\citenamefont {Kumar},
  \citenamefont {Panda},\ and\ \citenamefont {Sen}}]{Kumar:2013oda}%
  \BibitemOpen
  \bibfield  {author} {\bibinfo {author} {\bibfnamefont {S.}~\bibnamefont
  {Kumar}}, \bibinfo {author} {\bibfnamefont {S.}~\bibnamefont {Panda}}, \ and\
  \bibinfo {author} {\bibfnamefont {A.~A.}\ \bibnamefont {Sen}},\ }\href
  {\doibase 10.1088/0264-9381/30/15/155011} {\bibfield  {journal} {\bibinfo
  {journal} {Class. Quant. Grav.}\ }\textbf {\bibinfo {volume} {30}},\ \bibinfo
  {pages} {155011} (\bibinfo {year} {2013})},\ \Eprint
  {http://arxiv.org/abs/1302.1331} {arXiv:1302.1331 [astro-ph.CO]} \BibitemShut
  {NoStop}%
\end{thebibliography}%

\end{document}